\documentclass[11pt]{article}

\usepackage[usenames,dvipsnames,table]{xcolor}

\usepackage{framed}

\definecolor{shadecolor}{rgb}{0.90,0.90,0.90}

\usepackage{changepage}

\pdfoutput=1 
\usepackage{jheppub} 

\usepackage{graphicx}
\usepackage{amsmath, amssymb}

\usepackage{manfnt}

\newcommand{\be}{\begin{eqnarray}}
\newcommand{\ee}{\end{eqnarray}}
\newcommand{\bea}{\begin{eqnarray}}
\newcommand{\eea}{\end{eqnarray}}

\newcommand{\bn}{\begin{enumerate}}
\newcommand{\en}{\end{enumerate}}

\def\Tr{\mathop{\mathrm{Tr}}\nolimits}

\def\SU{\mathrm{SU}}

\title{SQCD and pairs of pants}

\author{Shlomo S. Razamat and Evyatar Sabag}

\affiliation{Department of Physics, Technion, Haifa, 32000, Israel}

\emailAdd{razamat@physics.technion.ac.il}
\emailAdd{sevyatar@campus.technion.ac.il}

\abstract{ We show that the $4d$ ${\cal N}=1$ $SU(3)$ $N_f=6$  SQCD is   the model obtained when compactifying  the rank one E-string theory on a three punctured sphere (a trinion) with a particular value of flux. 
The $SU(6)\times SU(6)\times U(1)$ global symmetry of the theory, when  decomposed into the $SU(2)^3\times U(1)^3\times SU(6)$ subgroup, corresponds to the three $SU(2)$ symmetries associated to the three punctures and the $U(1)^3 \times SU(6)$  subgroup of the $E_8$ symmetry of the E-string theory.
All the puncture symmetries are manifest in the UV and thus we can construct ordinary Lagrangians flowing in the IR to any compactification of the E-string theory.
We generalize this claim and argue that the ${\cal N}=1$ $SU(N+2)$ SQCD in the middle of the conformal window, $N_f=2N+4$, is the theory obtained by compactifying the $6d$ minimal $(D_{N+3},D_{N+3})$ conformal matter SCFT on a sphere with two maximal $SU(N+1)$ punctures, one minimal $SU(2)$ puncture, and with a particular value of flux. 
The $SU(2N+4)\times SU(2N+4)\times U(1)$ symmetry of the UV Lagrangian decomposes into $SU(N+1)^2\times SU(2)$ puncture symmetries and the $U(1)^3\times SU(2N+4)$ subgroup of the $SO(12+4N)$ symmetry group of the $6d$ SCFT.
The models constructed from the trinions exhibit a variety of interesting strong coupling effects. For example, one of the dualities arising geometrically from different pair-of-pants decompositions of a four punctured sphere  is an $SU(N+2)$ generalization of  the Intriligator-Pouliot duality of $SU(2)$ SQCD with $N_f=4$, which is a degenerate, $N=0$, instance of our discussion.}

\begin{document} 

\maketitle
\flushbottom

\section{Introduction} 

Interrelations between supersymmetric quantum field theories in different space-time dimensions often provide us with tools to organize, on one hand, our knowledge of the strongly coupled dynamics, and on the other hand to derive novel statements about said dynamics. 
One instance of these ideas, which sparked significant progress in understanding supersymmetric dynamics in recent years, is to consider $4d$ low energy theories of $6d$ $(1,0)$ SCFTs. 
These are obtained by putting the $6d$ SCFT on a Riemann surface \cite{Gaiotto:2009we} possibly with a choice of background fields, fluxes, for the global symmetry \cite{Chan:2000qc}.\footnote{This idea can be generalized to reductions to  lower dimensions, see {\it e.g} \cite{Dimofte:2011ju,Gadde:2013sca,Gukov:2018iiq}.} The geometric construction of the $4d$ models gives us a systematic way to understand IR dualities, conformal dualites, and emergence of symmetry, see \cite{Gaiotto:2015usa,Razamat:2016dpl,Kim:2017toz,Kim:2018bpg,Razamat:2018gro,Razamat:2019ukg,Pasquetti:2019hxf,Razamat:2018gbu,Sela:2019nqa} for some examples. 
 
 One of the main tasks to perform in such a geometric program is to compile a dictionary between $6d$ SCFTs supplemented with the geometric data of the compactification, and the $4d$ models.
 Typically the way this dictionary is compiled is by first deriving compactifications on simple building block surfaces, such as spheres with two (tubes) and three (trinions) punctures, and understanding the procedure of gluing these building blocks together. The latter involves gauging some of the global symmetries of the building blocks, possibly adding matter fields.
The question of finding the $4d$ models corresponding to two punctured sphere can be related to the problem of understanding reductions of the $6d$ model first to $5d$ and studying domain walls in the $5d$ low energy QFT \cite{Kim:2017toz,Gaiotto:2014ina}. However, a systematic general way to obtain the trinions is still lacking. One idea to do so was exploited in \cite{Razamat:2019mdt,Razamat:2019ukg} and involves considering $6d$ flows between different $6d$ SCFTs and the related flows between their compactifications. In particular it was shown that understanding compactifications on tubes of the $6d$ UV SCFT can lead to understandings of compactifications on surfaces with more punctures of the $6d$ IR SCFT.
 
 In this note we consider an entry in the dictionary between compactifications of $6d$ SCFTs and $4d$ QFTs, which although was obtained in
  a rather indirect way, has an intriguing structure  \cite{Razamat:2019vfd}. Our goal will be to exploit this structure and derive a Lagrangian description in $4d$ for
   compactifications of a particular SCFT, the E-string theory, and one family of its generalizations.  Let us first review the simple observation of \cite{Razamat:2019vfd} and then outline our results and the structure of this paper.
   
Let us consider the E-string theory on a closed Riemann surface of genus $g$ without turning on flux for abelian subgroups of its $E_8$ global symmetry.  The anomaly polynomial of this $6d$ theory is known \cite{Ohmori:2014pca,Ohmori:2014kda} and we can integrate it on the Riemann surface \cite{Benini:2009mz} to obtain a prediction for the anomaly polynolmial of the $4d$ low energy SQFT obtained in the compactification procedure. The conformal anomalies for compactification on a genus $g>1$ surface are then  \cite{Kim:2017toz},
\be\label{estringan}
a=\frac{75}{16} (g-1)\,,\qquad c= \frac{43}8 (g-1)\,.
\ee 
We can search for a $4d$ QFT which has these anomalies.

\begin{figure}[htbp]
	\centering
  	\includegraphics[scale=0.36]{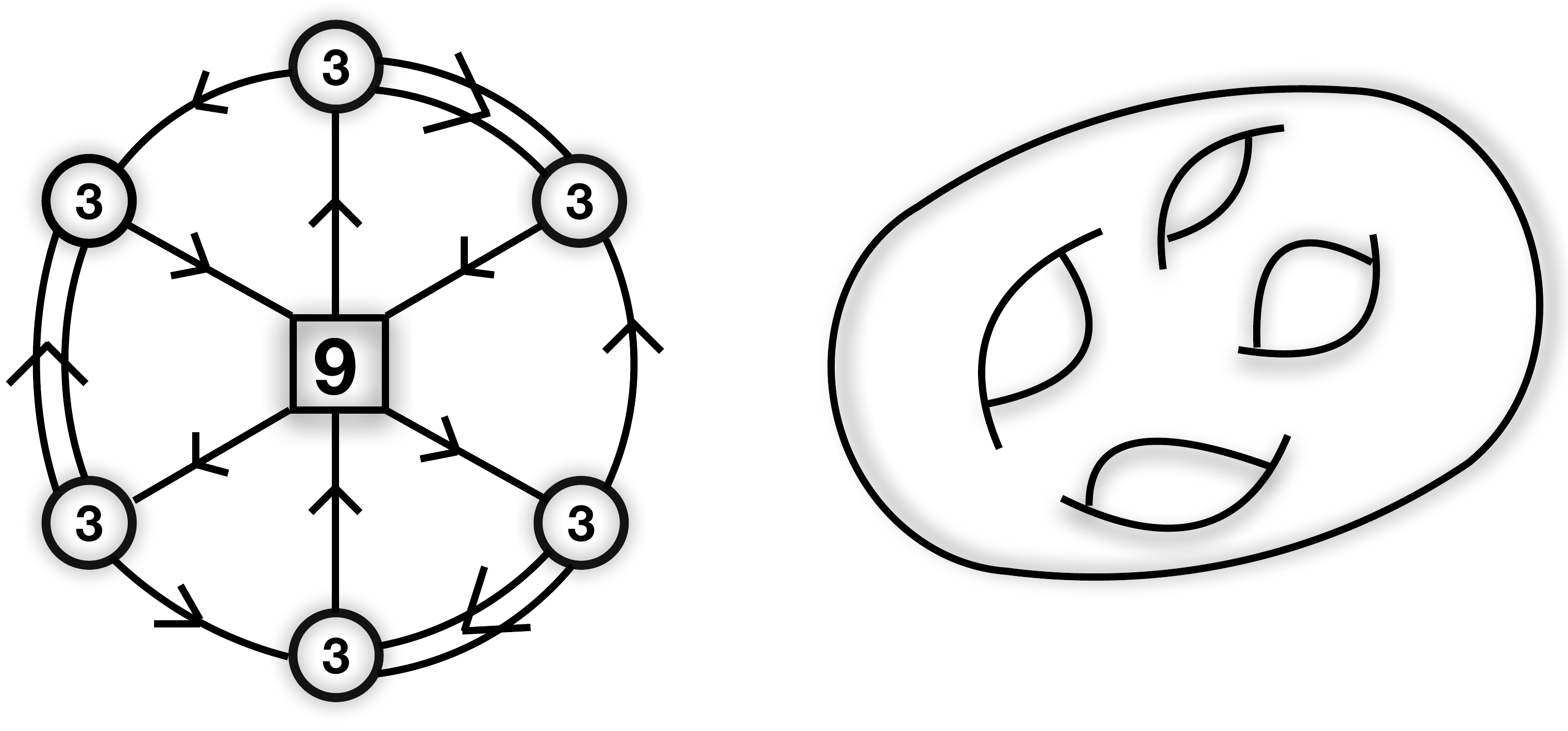}
    \caption{The {\it wheel} quiver theory corresponding to compactification of the E-string on genus $g$ surface with zero flux. The R-charges of all of the fields are $2/3$ and we have superpotential terms for all of the triangles as well as baryonic superpotentials for the edges of the circles. We can also turn on baryonic superpotentials for the spikes but these will further break the rank of the global symmetry group. The $SU(9)$ symmetry manifestly appearing in the quiver is a maximal subgroup of $E_8$. On some locus of the conformal manifold it is conjectured that the Cartan of $SU(9)$ enhances to the $E_8$ symmetry of the E-string. The model corresponding to the compactification on genus $g$ surface with zero flux has $2g-2$ $SU(3)$ gauge nodes.}
    \label{F:wheel}
\end{figure}  

In \cite{Razamat:2019vfd} the possibility of the pertinent QFT being a conformal weakly coupled SCFT was entertained. If this is the case then the conformal anomalies can be directly related to the dimension of the gauge symmetry group, $\text{dim}\,{\frak G}$, and the dimension of the representation of the matter fields, $\text{dim}\,{\frak R}$, (as the superconformal R-charges are the free ones),
\be\label{genan}
a=\frac3{16}\text{dim}\,{\frak G}+\frac1{48}\text{dim}\,{\frak R}\,,\qquad\qquad c=\frac1{8}\text{dim}\,{\frak G}+\frac1{24}\text{dim}\,{\frak R}\,.
\ee 
Using this relation and \eqref{estringan} one can quickly obtain that the model in $4d$, if indeed it has a conformal Lagrangian description, should have $\text{dim}\,{\frak R}=81(g-1)$ and
 $\text{dim}\,{\frak G}=16(g-1)$. Analyzing the low genus cases one can quickly arrive at the conclusion that a conformal gauge theory with $2g-2$ $SU(3)$ gauge groups and matter in (bi)fundamental representation can fit the bill. The precise model is depicted in Figure \ref{F:wheel}.
This theory is indeed conformal as each $SU(3)$ node has nine flavors and one can turn on various baryonic superpotentials, see \cite{Leigh:1995ep,Green:2010da}. The $SU(9)$ global symmetry depicted in the quiver is a maximal subgroup of $E_8$ and the claim of \cite{Razamat:2019vfd} is that on some locus of the conformal manifold of this model the Cartan of the $SU(9)$ enhances to the full $E_8$ group. One can check this claim by computing the supersymmetric index \cite{Kinney:2005ej,Romelsberger:2005eg,Dolan:2008qi,Rastelli:2016tbz} and various 't Hooft anomalies.  Moreover the dimension of the conformal manifold is given for a generic genus by $(3g-3)+248(g-1)$ with the former factor corresponding to the complex structure moduli of the compactification surface and the latter to the space of flat connections on the said surface \cite{Razamat:2016dpl,Morrison:2016nrt}.
 
The question we then want to ask is how one can understand the quiver of Figure \ref{F:wheel} as a composition of theories corresponding to trinion building block geometries. In what follows we will find a surprisingly simple answer to this question with the basic ingredient being,
\begin{shaded}
\be
&&\text{Trinion with }\, \frac12 \, \text{unit of flux breaking } \, E_8 \, \text{to }\, E_7\times U(1) \;\; \to \;\; {\cal N}=1\,\; \text{SU(3) } \,\; N_f=6 \,\;  \text{SQCD}\,\nonumber  
\ee
\end{shaded} 
Gluing such trinions by gauging $SU(2)$ subgroups of the $SU(6)$ symmetry rotating the fundamentals and adding certain superpotentials we can obtain the model in Figure \ref{F:wheel}. In what follows we will detail how this claim comes about. Let us stress that a lot is known by now about compactifications of E-string.
In \cite{Kim:2017toz} the tube models were derived, and a construction of a trinion model was presented. The latter involved gauging of IR emergent symmetries.
In \cite{Razamat:2019ukg} a completely Lagrangian construction of a trinion was introduced, however one of the puncture symmetries was emergent in the IR so to build theories for arbitrary surfaces one has to gauge these emergent symmetries. The construction we detail here does not involve gauging of IR emergent symmetries and thus can be described by conventional Lagrangians which flow in the IR to the required models, and that might be strongly coupled SCFTs. As we will discuss, although the theories we will construct have regular Lagrangians with manifest symmetries, their dynamics is typically intricate involving various dangerously irrelevant deformations in the UV. 
Using the above basic building block one can construct a variety of  SCFTs exhibiting interesting properties such as conformal dualities, IR dualities, and emergence of symmetry.
  
  We can generalize the above statement in two ways.  First, the rank one E-string theory, can be viewed as the low energy model of a single M5 brane probing a $D_4$ singularity. This point of view has a generalization to models residing on a single M5 brane probing a $D_{N+3}$ singularity. These models are typically called minimal $(D_{N+3},D_{N+3})$ conformal matter SCFTs \cite{DelZotto:2014hpa}, and we will  refer to them as minimal $D_{N+3}$ conformal matter for brevity. The global non R-symmetry of these models is $SO(4N+12)$, which enhances to $E_8$ for $N=1$. We will argue that trinions here with two maximal punctures and one minimal have simple properties and in particular,
  \begin{shaded}
\be
&\text{Trinion with}\, \frac12 \, \text{unit of flux breaking} &\,\nonumber\\
&SO(4N+12) \, \text{ to }\, SO(4N+8)\times SU(2) \times U(1) \, & \to \;\;\;\; \, {\cal N}=1\,\; \text{SU(N+2)} \,\; N_f=2N+4 \,\;  \text{SQCD}\,\nonumber  
\ee
\end{shaded} 
Moreover, the second generalization of the above statement is that we will argue  that trinions with some flux, one maximal, one minimal, and one generic puncture are related to ${\cal N}=1$ $SU(N+2)$ SQCD with  $N_f\leq2N+4$.
 
The paper is organized as follows. In section \ref{sec:E} we discuss our basic claim about E-string compactifications detailed above. In section \ref{sec:D} we generalize the statements to compactifications of the minimal $D_{N+3}$ conformal matter.   We also give a geometric interpretation of  Seiberg duality \cite{Seiberg:1994pq} of SQCD in the middle of conformal window and generalize the Intriligator-Pouliot duality \cite{Intriligator:1995ne} in this section.
In section \ref{sec:A} we discuss our results. 
In appendix \ref{app:oldtrinion} we detail how the trinion of the current paper is related to the one of  \cite{Razamat:2019ukg}.

\section{E-string trinion and compactifications}\label{sec:E}

\subsubsection*{The trinion}
\begin{figure}[htbp]
	\centering
	\includegraphics[scale=0.38]{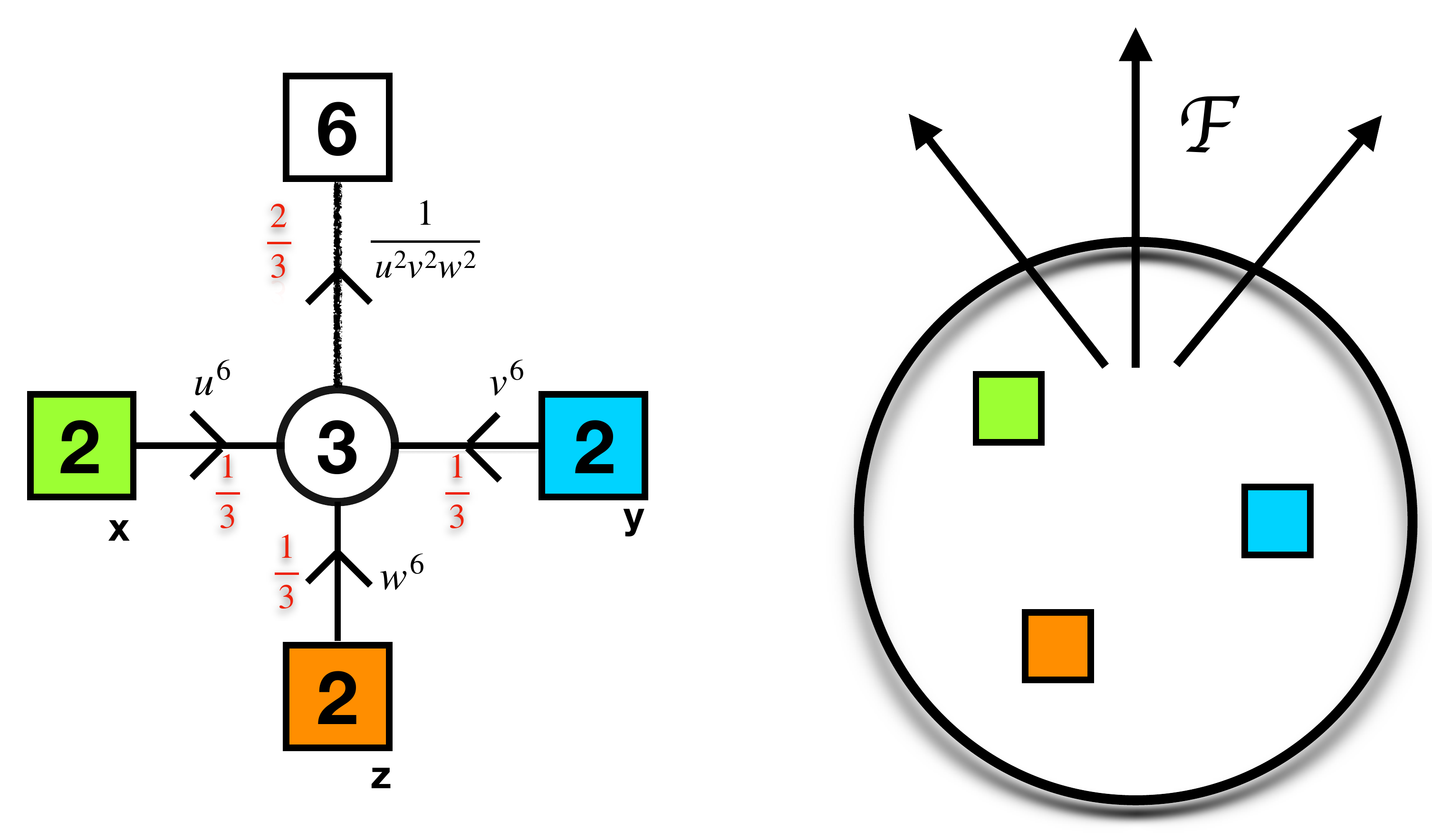}
    \caption{A quiver description of the suggested E-string trinion. The choice of R-charges, here and in all other figures in the paper, is denoted in red. This choice is not the superconformal one, where all fields have R-charge half, but is the one inherited from six dimensions. There are no superpotential terms. The three maximal punctures are of different ``color'' reflected in a different pattern of charges under the Cartan of the $E_8$ of the corresponding moment map operators. The charges under $u$, $v$, and $w$ are motivated by not having fractional powers in the higher $N$ generalizations.}
    \label{F:trinionE}
\end{figure}

We consider ${\cal N}=1$ $SU(3)$ $N_f=6$  SQCD and claim that it should be associated with compactification of the E-string theory on a three punctured sphere with half a unit of flux braking $E_8$ global symmetry to $U(1)\times E_7$.  The model is depicted in Figure \ref{F:trinionE}. We split the $6$ fundamental chiral superfields  into three pairs and correspondingly decompose the $SU(6)$ symmetry rotating them into $SU(2)^3\times U(1)^2$. Each $SU(2)$ factor is associated to a puncture. In addition we have the two $U(1)$ symmetries arising in the decomposition, the baryonic $U(1)$, and the $SU(6)$ rotating the antifundamentals. This gives us a rank eight symmetry which is embedded in $E_8$ as
\be
SU(6)\times U(1)^3 \subset SU(6)\times SU(3)\times U(1) \subset E_7\times U(1) \subset E_8\,.
\ee  
The flux is in the last $U(1)$.  

We assign an R-charge $1/3$ to the fundamentals and $2/3$ to the anti-fundamentals. Note that this is a non anomalous symmetry and it corresponds to the R-charge directly inherited from $6d$ as we will soon discuss. Under the superconformal R-symmetry all chiral superfields are assigned R-charge $1/2$, but we will refrain from using it.
Each puncture has an octet of operators which transform in the fundamental representation of the puncture $SU(2)$ symmetry.
 These have R-charge $+1$ and six of them are composed of mesons while two are baryons. We will refer to these operators as ``moment maps''.
 For each puncture their charges are given by the following,
\be
&&M_u={\bf 2}_x\;\otimes\; \left({\bf 6}_{u^4/v^2w^2}\oplus {\bf 1}_{u^6 v^{12}}\oplus {\bf 1}_{u^6 w^{12}}\right)\,,\nonumber\\
&&M_v={\bf 2}_y\;\otimes\; \left({\bf 6}_{v^4/u^2w^2}\oplus {\bf 1}_{v^6 u^{12}}\oplus {\bf 1}_{v^6 w^{12}}\right)\,,\\
&&M_w={\bf 2}_z\;\otimes\; \left({\bf 6}_{w^4/u^2v^2}\oplus {\bf 1}_{w^6 u^{12}}\oplus {\bf 1}_{w^6 v^{12}}\right)\,.\nonumber
\ee 
Here we denote by $\{x,y,z\}$ the Cartans of the three $SU(2)$ puncture symmetries and ${\bf 6}$ is the fundamental of the $SU(6)$ symmetry. Here and throughout the paper we will encode the charges of operators under various symmetries in powers of fugacities associated to them.
The operators in the ${\bf 6}$ of the $SU(6)$ symmetry come from mesons and the singlets are baryons. For example to construct the baryon charged  ${\bf 1}_{u^6 v^{12}}$ we consider the antisymmetric square of the fundamental quark charged under $SU(2)_y$ which gives us a singlet of $SU(2)_y$ transforming in the antifundamental representation of the gauge $SU(3)$. Contracting it with the fundamental of $SU(2)_x$ gives us a singlet with the given charges.\footnote{Note that there are more baryonic operators but these are charged under more than one puncture symmetry and will not play a role in our discussion. }
 The three punctures have moment map operators with different pattern of charges under the  various symmetries. Thus these should be considered as being of different types, {\it colors} in the nomenclature of \cite{Kim:2017toz}.
 
 We remind the reader that one can think about the puncture symmetry by considering the given $6d$ SCFT compactified on a circle, possibly with holonomies for the global symmetry group. In preferable situations one lands on an effective low energy $5d$ description as a gauge theory. Further compactifying the model to $4d$ on a segment choosing supersymmetry preserving boundary conditions, the $5d$ gauge symmetry becomes a global symmetry associated to the boundary (the puncture). Moreover, we expect to get operators in $4d$ charged under the puncture symmetry which come from $5d$ fields with Neumann boundary conditions. In the case of $A_{N-1}$ $(2,0)$ theory the $5d$ model is maximally supersymmetric SYM with gauge group $SU(N)$ and the relevant fields surviving as operators in $4d$ give us an adjoint chiral operator. In $4d$ the theories might have   ${\cal N}=2$ supersymmetry and then these adjoint operators are the moment maps. For E-string the relevant $5d$ model is $SU(2)$ ${\cal N}=1$ model with eight fundamentals. Thus, we have an $SU(2)$ symmetry associated to the puncture and an octet of fundamental operators, to which by analogy with the $(2,0)$ case we refer to as the ``moment maps''.
  
Let us also mention some of the anomalies of the model,
\be
&&\Tr\, R =  -10\,,\qquad \Tr\, R^3 =  2\,,\qquad \Tr\, R\, SU(2)_{x,y,z}^2 =  -1\,.
\ee These are indeed the expected anomalies of the compactification of E-string on  three punctured sphere where $R$ stands for the Cartan generator of the $6d$ $SU(2)_R$ R-symmetry \cite{Kim:2017toz,Razamat:2019ukg,Pasquetti:2019hxf}. These anomalies are independent of the value of the flux for the global symmetry. Next, we will derive the flux of the trinion by gluing trinions together to form closed Riemann surfaces.

\subsection{Gluings}

\subsubsection*{S-gluing}
Next we consider combining the trinions into a closed Riemann surface. First we consider the S-gluing \cite{Kim:2017toz}.\footnote{For S-gluing in other contexts see \cite{Bah:2012dg,Gaiotto:2015usa,Hanany:2015pfa,Razamat:2016dpl}.}
This is the procedure of gluing two punctures together by gauging the puncture $SU(2)$ symmetry and coupling the moment maps of the two punctures, the octets $M$ and $M'$, through the superpotential,
\be\label{Esuperpotential}
W= \sum_{i=1}^8  M_i M'_i\,.
\ee 
In particular if we take two identical trinions and S-glue them together the symmetries of one are identified with the conjugation of the other. This in particular implies that the flux of the combined surface is zero. Note that our R-charge is not anomalous under this gluing as the $SU(2)$ gauging here have three flavors.  As the moment maps are mesons and some of the  baryons, the superpotential has quartic and sextic terms. 

We are after a map between compactifications of a $6d$ model and $4d$ effective field theory descriptions. The effective $4d$ theories we construct are conjecturally completed in the UV by the six dimensional SCFT, with the compact geometry serving as the relevant deformation triggering the RG flow across dimensions.\footnote{For some general classes of examples of flow analysis between theories in lower dimensions one can consult~\cite{Aharony:2013dha,Aharony:2017adm}.}
We might wonder whether the models are UV complete as $4d$ models and what is their behavior in the IR. The $4d$ model might have a variety of IR behaviors. In particular it might be an interacting SCFT, it might have some sectors decoupling in the IR (see for example \cite{Chacaltana:2010ks,Kim:2017toz,Kim:2018bpg,Kim:2018lfo,Bah:2017gph,DelZotto:2015rca}), and it might have IR free gauge groups \cite{Ohmori:2015pia}. 
There are known cases which can serve as an example for all these behaviors.
In explicit computations performed in various setups till now one can deduce the vague impression that typically compactifications on high enough genus and with high enough flux lead to interacting SCFTs in the IR. With this in mind let us first analyze the dynamics of this gluing.  As we will see the dynamics is not trivial.\footnote{ Related to this, let us mention here that the anomalies of the symmetries inherited from $6d$ of the putative $4d$ models obtained in compactification should match the $6d$ predictions, even if the theory has accidental symmetries in the IR and/or the gauge groups are free. This is true assuming we could identify the $6d$  symmetries correctly. In that regard, later on we will match the conformal anomalies $a$ and $c$ obtained using the $a$-maximization procedure \cite{Intriligator:2003jj}. We will not analyze the dynamics in all the cases in detail. These anomaly matchings should be viewed as an agreement of  certain combinations of 't Hooft anomalies when the dynamics is not analyzed.} 

\begin{figure}[t]
	\centering
  	\includegraphics[scale=0.31]{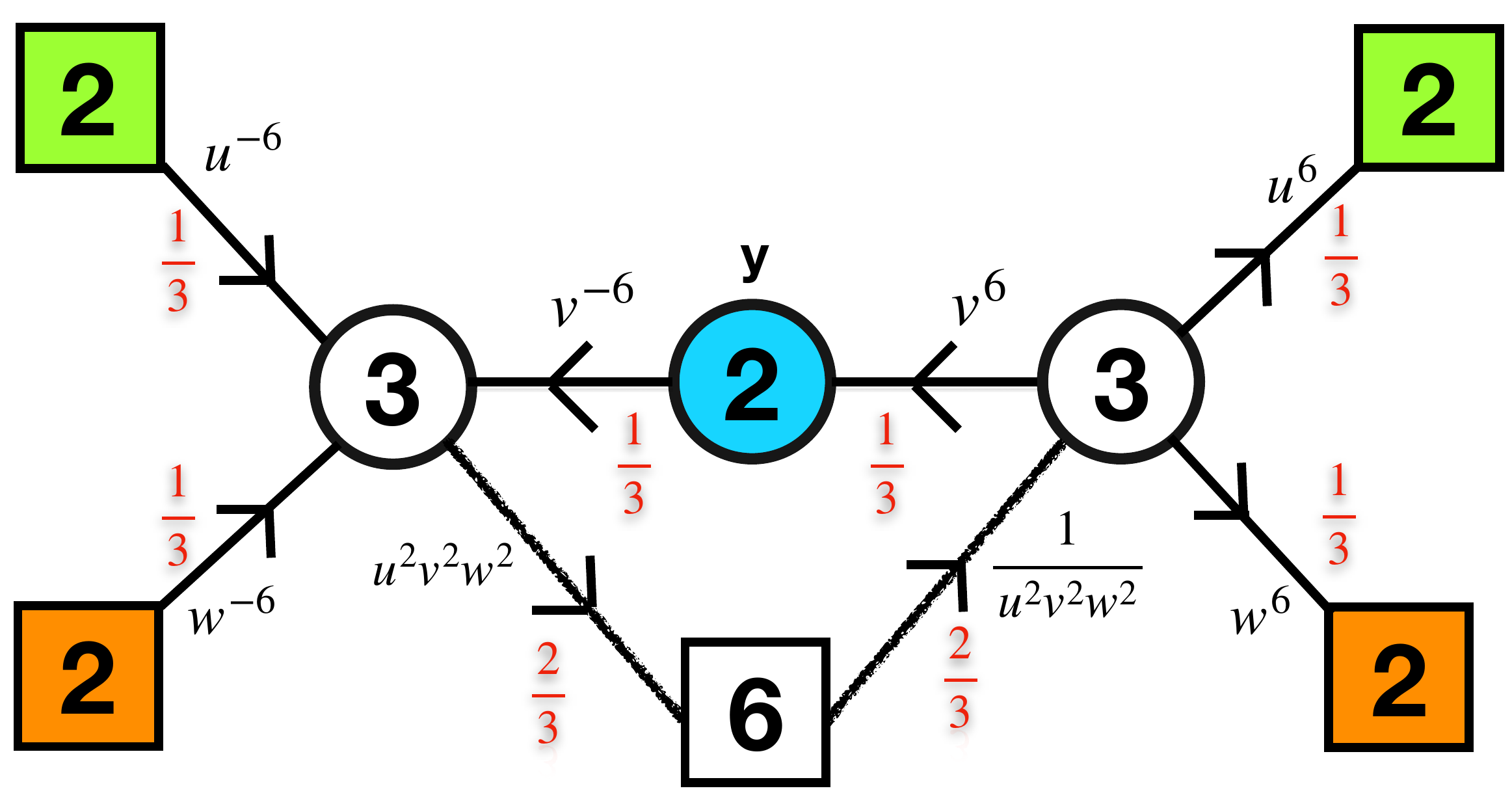}\;\;\; 	\includegraphics[scale=0.31]{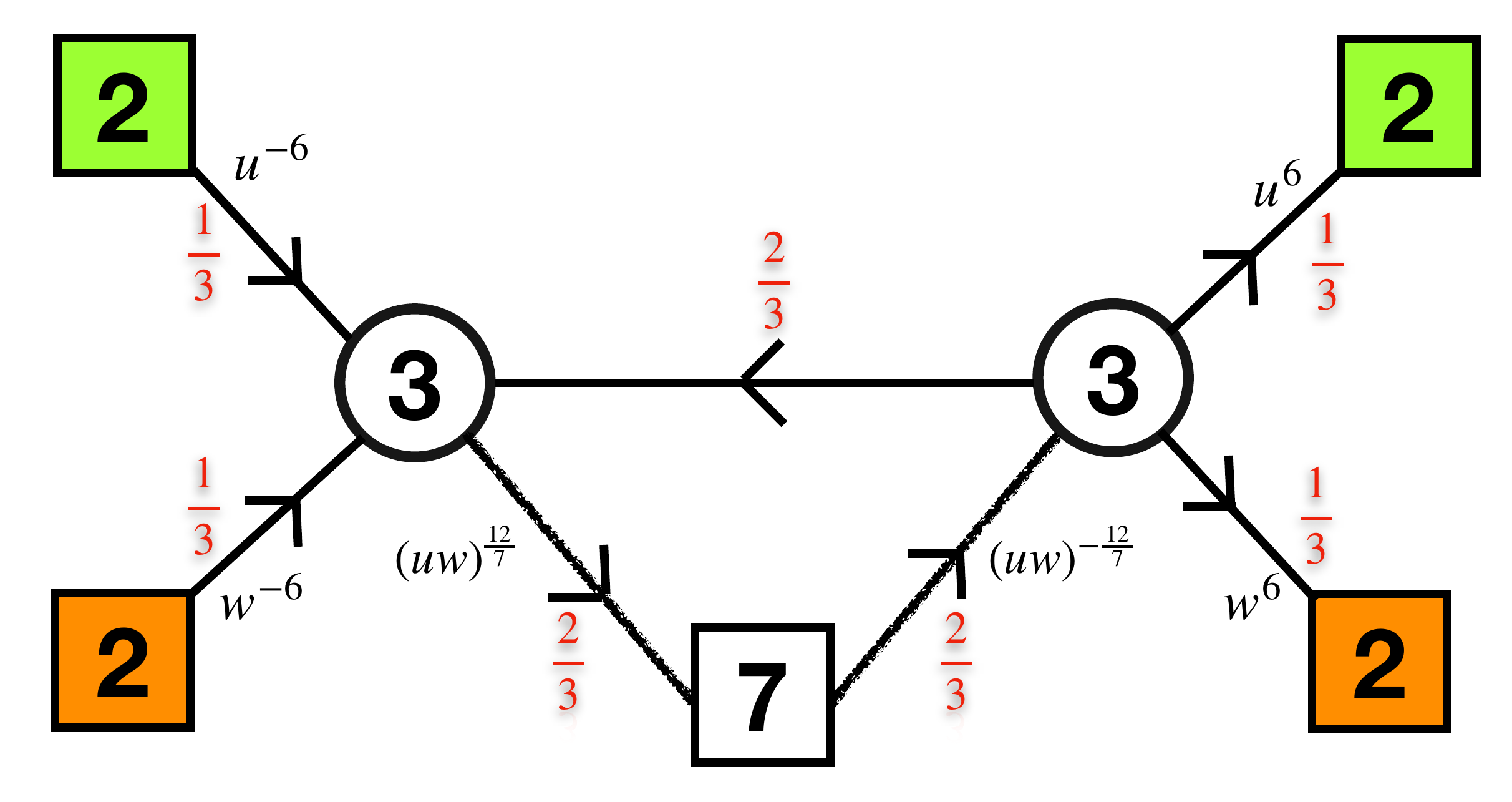}
    \caption{S-gluing of the two trinions. There are quartic and sextic superpotentials locking the various symmetries as indicated. On the left we have the two trinions glued and on the right the same model after performing Seiberg duality on the $SU(2)$ node.  The Seiberg dual  is a WZ model built from mesons and baryons of the $SU(2)$ gauge group. The former will constitute the bifundamental between the two $SU(3)$ nodes, while the latter add an (anti)fundamental to each one of the $SU(3)$ nodes.}
    \label{F:sglued}
\end{figure}

In our gluing we have various interactions: several types of superpotential terms and gauging of a global symmetry, see Figure \ref{F:sglued}.
A way to analyze the dynamics in such situations is to turn on deformations one by one. 
 One can flow to an infrared fixed point with a single relevant deformation or move on the conformal manifold with a single exactly marginal deformation. Once the deformation was performed, analyze the relevance of deformations at the fixed point, and turn on the next deformation. Deformations which are irrelevant at a certain step of the flow might become relevant or exactly marginal after performing several deformations. In such a situation they are called {\it dangerously irrelevant}. This analysis is important to understand what are the properties of the infrared fixed point of the theory we are after.
In particular this is crucial for discussing gauging of symmetries: the beta function of certain gauge interaction might be IR free, and thus such a gauging might appear to be not UV complete at some stage of the flow. This might change at a later stage, as we are in a new strongly coupled fixed point. Performing the analysis of the beta function at this point one might discover that it is now exactly marginal or asymptotically free making the construction UV complete. In supersymmetric theories this analysis is greatly simplified since the beta function of a gauge group $G$ coupling is related to a simple 't Hooft anomaly, $\Tr \, R\, G^2$, with $R$ being the superconformal R-symmetry of the fixed point at which we gauge the symmetry $G$ weakly.
Positive sign of this 't Hooft anomaly implies gauging is asymptotically free, negative that it is IR free, and zero that it is marginal (see {\it e.g.} \cite{Benini:2009mz}). The superconformal R-symmetry is fixed by $a$-maximization \cite{Intriligator:2003jj} which  presumes that we do flow to an SCFT, that there are no accidental abelian symmetries, and in particular no decoupling sectors \cite{Kutasov:2003iy}.
Let us outline how this works  for the S-gluing.
 
The superconformal R-symmetry of both trinions sets an R-charge $+\frac12$ for all the chiral fields and thus the gauging is asymptotically free and the quartic superpotentials are marginal. In the following sequence of deformations we will turn on these interactions before sextic ones. 
Performing the gauging first will identify the $U(1)_v$ symmetry of one of the trinions with the conjugate of the other. This in turn will set the R-charge of the (anti)fundamentals not charged under the gauged $SU(2)$ to $8/15$ and the ones charged under gauged $SU(2)$ to $1/3$, making the quartic superpotential relevant.
Turning on the quartic superpotenial locks the  $SU(6)$ symmetries and $U(1)_{u v w}$ symmetries of the two theories. Computing the superconformal R-symmetry by a-maximization \cite{Intriligator:2003jj} with these interactions we obtain that now the $6d$ R-symmetry is the superconformal one.
Then in the IR the baryonic moment maps have R-charge $+1$, meaning that the superpotential we want to turn on is marginal. The charges of one of the baryonic superpotential terms is $u^{12}u'^{12}$ while the other has charge $w^{12}w'^{12}$, where the prime distinguishes the symmetries of the two trinions. The quartic superpotential we have turned on before, locks the four symmetries to have $(uu'ww')^{2}=1$, meaning that the two marginal operators we turn on have opposite charges. This in turn implies that turning both of them is an exactly marginal deformation \cite{Green:2010da}.\footnote{See \cite{Razamat:2020pra,Razamat:2020gcc} for many examples of group theoretic computations of exact marginality of operators.} We thus expect to arrive in the IR to an interacting SCFT pending the usual caveats of no accidental abelian symmetries in the IR.

Note that as the $SU(2)$ gauging has three flavors we can use Seiberg dual description \cite{Seiberg:1994pq} giving us just a WZ model of the mesons, providing a bifundamental of the two $SU(3)$ symmetries, and baryons, giving us an (anti)fundamental for each one of the $SU(3)$ gauge groups. See Figure \ref{F:sglued}. All in all the model is a combination of two $SU(3)$ SQCD models with seven flavors.

To construct a genus two surface we need to glue the remaining four punctures in pairs. Note that when we glue two punctures on the same surface their type should be the same. As we have seen above the puncture type is encoded in the pattern of charges of the moment map operators.
If this is not the case we necessarily break some of the global symmetry of the theory.\footnote{The breaking of symmetries in such cases can be often associated to some discrete twists/fluxes for which global structure of the full symmetry group, including the puncture, is important (see \cite{Kim:2017toz,Bah:2017gph}). We will refrain here from dealing with these issues. See however an example in appendix \ref{app:oldtrinion}.}
Now all the gaugings and superpotentials are marginal. Note that gluing the next pair of  punctures and using Seiberg duality we will obtain two copies of $SU(3)$ SQCD with eight flavors, which will correspond to a torus with two punctures. Gluing the final pair of punctures we obtain two copies of $SU(3)$ SQCD with nine flavors. See Figure \ref{F:wheel} with $g=2$.  This is precisely the theory claimed in \cite{Razamat:2019vfd} to correspond to genus two compactification of E-string with zero flux.

\begin{figure}[htbp]
	\centering
  	\includegraphics[scale=0.37]{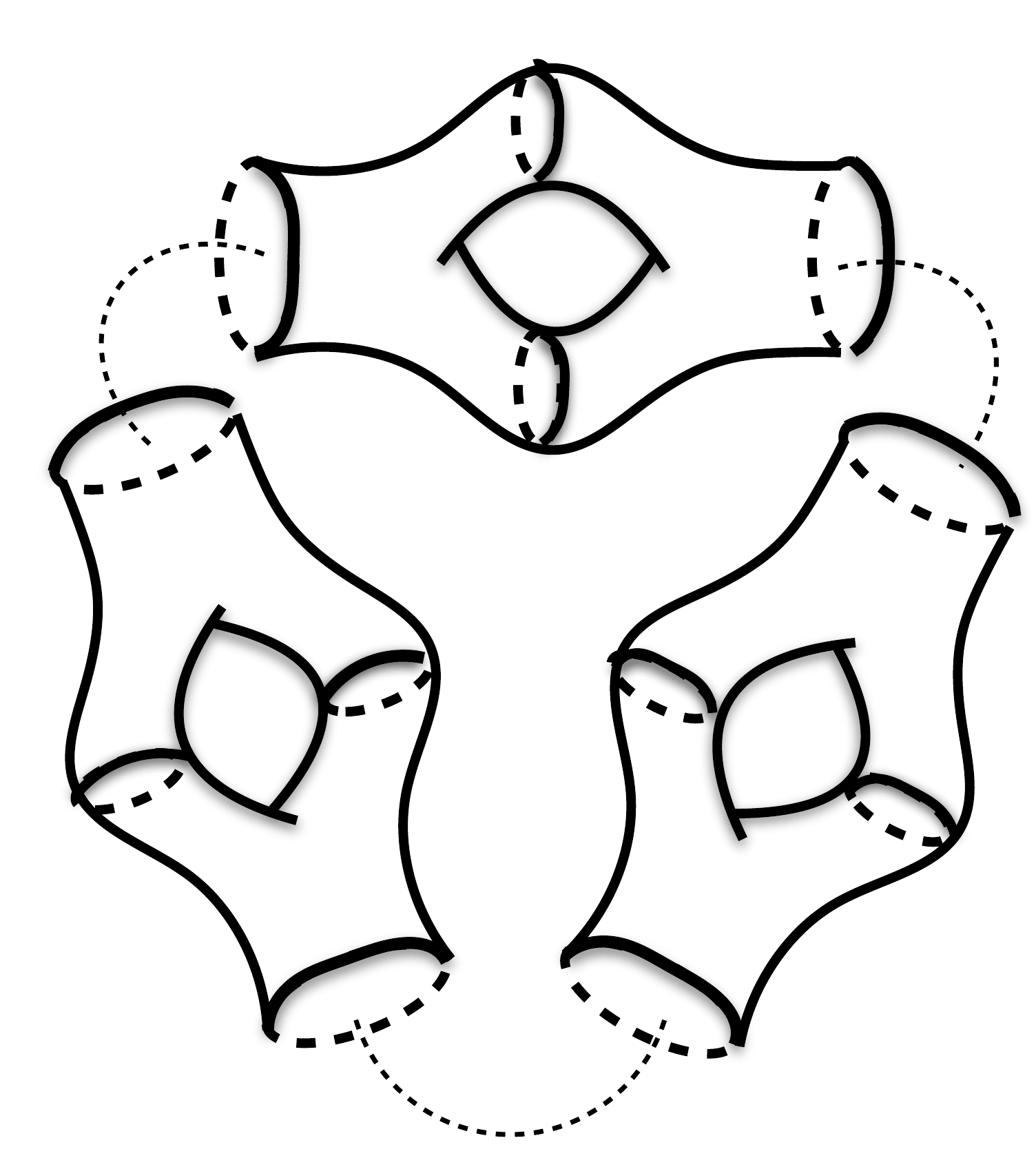}
    \caption{S-gluing trinions into the quiver of Figure \ref{F:wheel}. }
    \label{F:genus7}
\end{figure}

Instead of gluing the last two pairs of punctures together we can consider gluing $g-1$ two punctured tori in a circle to obtain a genus $g$ surface as depicted in Figure \ref{F:genus7}. Using the Seiberg dual frame of the $SU(2)$ gauge node with three flavors leads again to the same quiver as in Figure \ref{F:wheel} which is conjectured to be the rank one E-string theory compactified on genus $g$ surface with zero flux. Thus we recover the claim of \cite{Razamat:2019vfd}. 
Moreover the pair of pants decomposition of the surface should not matter and one can consider any decomposition resulting in same anomalies and same superconformal index. This implies certain dualities, we will comment on some of these in the next sub-section.

\subsubsection*{$\Phi$-gluing}
Next we consider $\Phi$-gluing \cite{Kim:2017toz}. This gluing includes gauging the puncture symmetry and identifying the moment map operators of the two punctures using the superpotential,

\be
W=\sum_{i=1}^8 \left(M_i-M_i'\right)\Phi_i\,,
\ee where $\Phi_i$ is an octet of fields in the fundamental representation of the gauged symmetry which are added to the model. Note here, that the symmetries of the glued punctures are identified without complex conjugation the flux of the glued surfaces is summed up. In particular this allows us to fix the flux of the trinion by comparing to predictions from six dimensional anomalies.

 \begin{figure}[htbp]
	\centering
  	\includegraphics[scale=0.40]{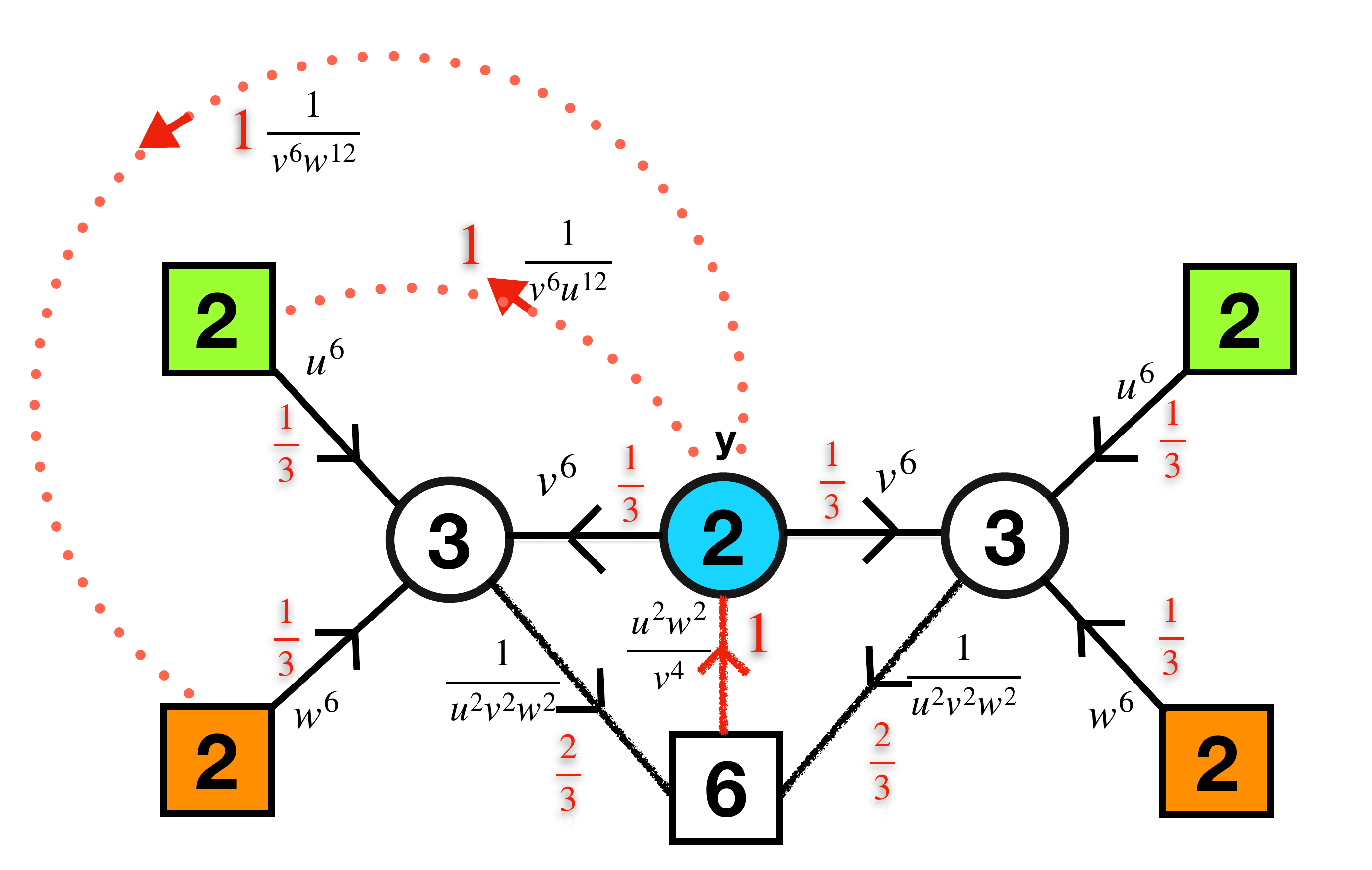}
    \caption{$\Phi$-gluing two trinions. The fields $\Phi_i$ flipping the moment maps are denoted in red. The dotted lines flips the baryonic operator with charges $v^6u^{12}$ and  $v^6w^{12}$. This fields are in the fundamental representation of the symmetry they emanate from and a singlet of the symmetry they point to.}
    \label{F:phigluing}
\end{figure}

The dynamics of this gluing is rather involved.
Note we added eight fields in the fundamental representation and the trinions contribute six more, so the $SU(2)$ gauging is IR free if done at the free UV fixed point. 
Also, as the superconformal R-symmetry of the chiral fields of the trinions is $+\frac12$, the gauging is IR free if done at the trinion fixed point. In particular the 't Hooft anomaly  $\Tr\, R\, SU(2)^2$ is negative at these fixed points.
 The superpotentials we turn on here are either cubic in fields (mesons coupled to six of the components of $\Phi_i$) or quartic (baryons coupled to two of the components of $\Phi_i$). The analysis of the dynamics here then proceeds in several steps, turning on a single relevant deformation in each step. Let us denote by $\Phi_{1,2}$ the two components of the chiral fields we add in the gluing which flip the baryonic moment map operators, and by $\Phi_{3\cdots8}$ the ones flipping the mesonic moment map operators.
 In step $(1)$ we take the two trinions, introduce the fields $\Phi_{3\cdots8}$, and turn on the cubic couplings, which are relevant deformations.
 These couplings lock the $SU(6)$ symmetries of the two trinions together and set $\left(\frac{wu}{v^2}\right)^2=\left(\frac{w'u'}{v'^2}\right)^2$. Flowing to the infrared we find a new superconformal R-symmetry which is obtained by a-maximization and involves mixing of the R-symmetry at the previous step with all the abelian symmetries. Next, we introduce at the new fixed point the fundamental fields $\Phi_{1,2}$ and compute $\Tr\, R\, SU(2)^2$ of the $SU(2)$ gauging using the new R-symmetry. It turns out that now it is in fact asymptotically free, and concretely $\Tr\, R\, SU(2)^2 =0.170705>0$.
 Thus, at step $(2)$ we gauge the symmetry with the two additional fundamental fields and flow again.
  The gauging  breaks one combination of the $U(1)$ symmetries making it anomalous. One way to think about it is that the requirement of the mixed gauge anomalies to vanish fixes the $U(1)$ symmetry of $\Phi_{1,2}$ in terms of the other symmetries such that the product of their fugacities equals $(wuv)^{-6}(w'u'v')^{-6}$. After this gauging, employing a-maximization sets the superpotential terms coupling the baryons to the $\Phi_{1,2}$ to have R-charge $2$, making these deformations marginal. Turning the quartic superpotential involving baryons at step $(3)$ we can construct an exactly marginal deformation and the primed and un-primed symmetries are locked on each other. The superconformal R-symmetry in the end is,
\be
R'=R+0.01938\,(q_u+q_w)+0.0372\,q_v\,.
\ee 
Here $q_{u,v,w}$ are charges under $U(1)_{u,v,w}$.
Thus we can conjecture that the model flows to an interacting SCFT. Note that the dynamics is rather involved and from the
 perspective of the UV Lagrangian has several dangerously irrelevant deformations.  

We can continue and $\Phi$-glue the remaining punctures in pairs. In particular by gluing another pair we will obtain a torus with two punctures and gluing all three punctures we obtain a genus two surface. 
For each such gluing we find that the same sequence of relevant deformations as above is required: cubic superpotentials, gauging, and then quartic superpotentials.
Gluing all three pairs of punctures together in such a manner we obtain that the superconformal R-symmetry at the fixed point is,
\be\label{mixe7}
R^{g=2}=R+\frac1{54}(\sqrt{5}-1)( q_u+q_v+q_w)\,,
\ee 
and the conformal anomalies are,
\be
 a=\frac{5 \sqrt{5}}{4}+\frac{47}{16}\,,\qquad c= \frac{3 \sqrt{5}}2+\frac{27}8\,.
\ee 
This agrees with the $4d$ anomalies obtained by integrating the anomaly polynomial of the E-string SCFT on genus $2$ surface with $1$ unit of flux in a $U(1)$ subgroup breaking $E_8$ to $E_7\times U(1)$ \cite{Kim:2017toz}. Thus each trinion is to be associated to half a unit of this flux. The $U(1)$ with the flux is the diagonal combination of $u$, $v$, and $w$ as can be understood from the mixing in \eqref{mixe7}. Then the remaining two $U(1)$s and the $SU(6)$ parametrize $SU(3)\times SU(6)$ which is a maximal subgroup of $E_7$. The branching rules are as follows,
 \be
&& E_8\; \to\; E_7\times SU(2)_{u^6v^6w^6} \;\to \; SU(6)\times SU(3)_{u^8/(w^4v^4),v^8/(w^4u^4) }\times SU(2)_{u^6v^6w^6}\,,\nonumber\\
&&{\bf 248}_{E_8}\to {\bf 133}_{E_7}\oplus {\bf 3}_{SU(2)}\oplus \left({\bf 2}_{SU(2)} \otimes {\bf 56}_{E_7}\right)\,,\\
&&{\bf 133}_{E_7}\to {\bf 35}_{SU(6)}\oplus {\bf 8}_{SU(3)}\oplus \left({\bf 15}_{SU(6)}\otimes {\bf 3}_{SU(3)}\right)\oplus \left(\overline{\bf 15}_{SU(6)}\otimes \overline{\bf 3}_{SU(3)}\right)\,,\nonumber\\
&&{\bf 56}_{E_7}\to \left({\bf 6}_{SU(6)}\otimes \overline{\bf 3}_{SU(3)}\right)\oplus \left(\overline{\bf 6}_{SU(6)}\otimes {\bf 3}_{SU(3)}\right)\oplus {\bf 20}_{SU(6)}\,.\nonumber
 \ee 
 Note that although the flux of the trinion preserves the $E_7\times U(1)$ subgroup of $E_8$, the trinion has a smaller symmetry as it has punctures. Punctures are defined through a $5d$ construction and involve boundary conditions for variuos fields, choice of which breaks the symmetry. The apperent symmetry of a theory is thus the intersection of symmetries preserved by the flux and the punctures \cite{Razamat:2016dpl,Kim:2017toz}. Once the trinions are glued into a closed surface we should observe the full symmetry preserved by the flux.
   For example computing the supersymmetric index \cite{Kinney:2005ej,Romelsberger:2005eg,Dolan:2008qi,Rastelli:2016tbz} of the genus two theory obtained by S-gluing the two trinions 
leads to the following leading order expansion in superconformal fugacities,
\be\label{indexgen2}
1+(4+{\bf 248}_{E_8} )\, qp+\cdots\,,
\ee 
with ${\bf 248}_{E_8}$ decomposed as above into the symmetries of the UV Lagrangian. In particular we see the $E_8$ symmetry emerging in the IR.
  
\subsection{Dualities and closing punctures}
 
 If the models we discussed are to be associated to punctured Riemann surfaces they need to satisfy duality properties associated to different pair of pants decompositions. Let us discuss the duality properties of the two types of gluings.
 
 \subsubsection*{S-gluing dualities}
 
\begin{figure}[htbp]
	\centering
  	\includegraphics[scale=0.27]{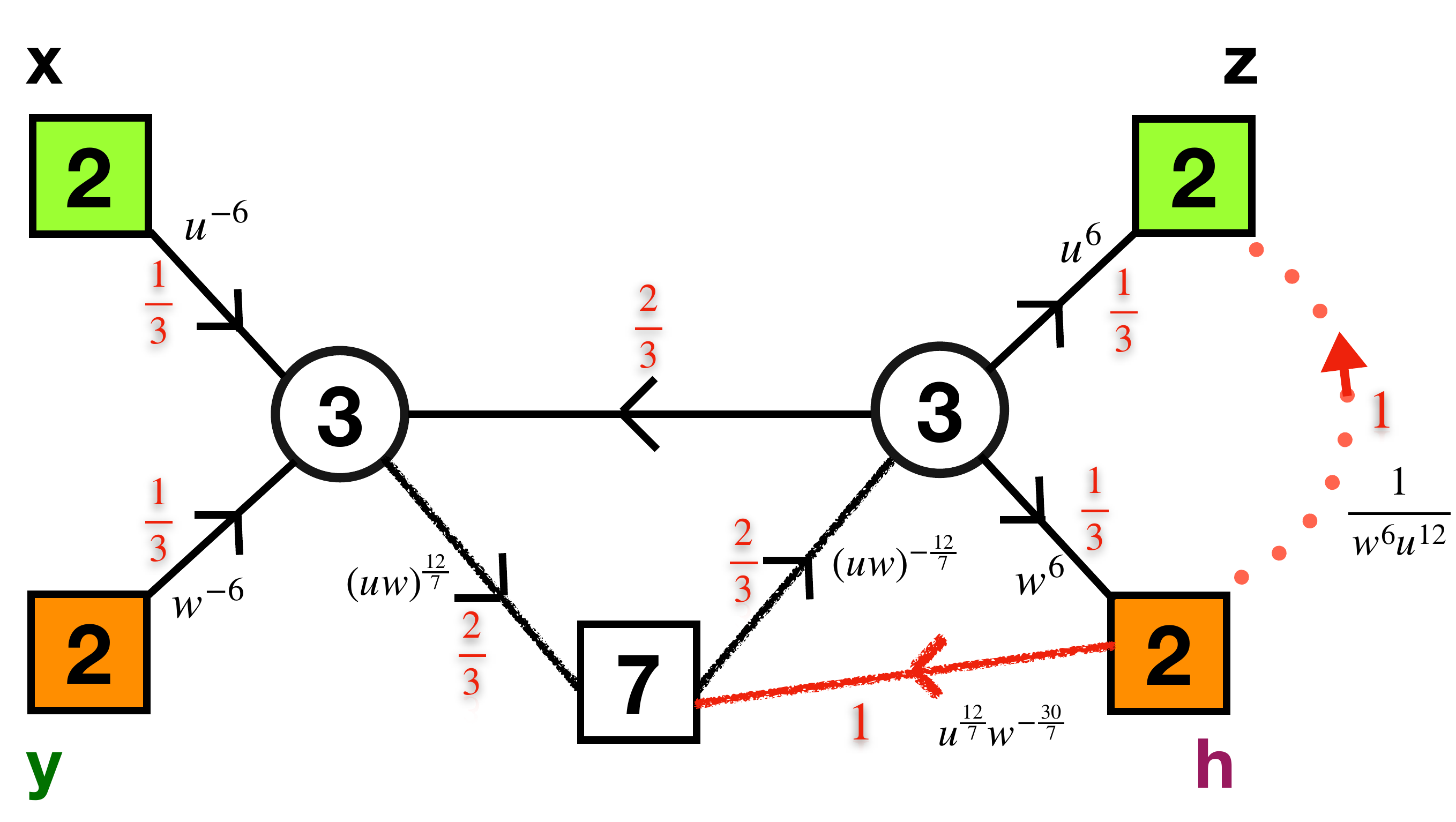}\; \includegraphics[scale=0.27]{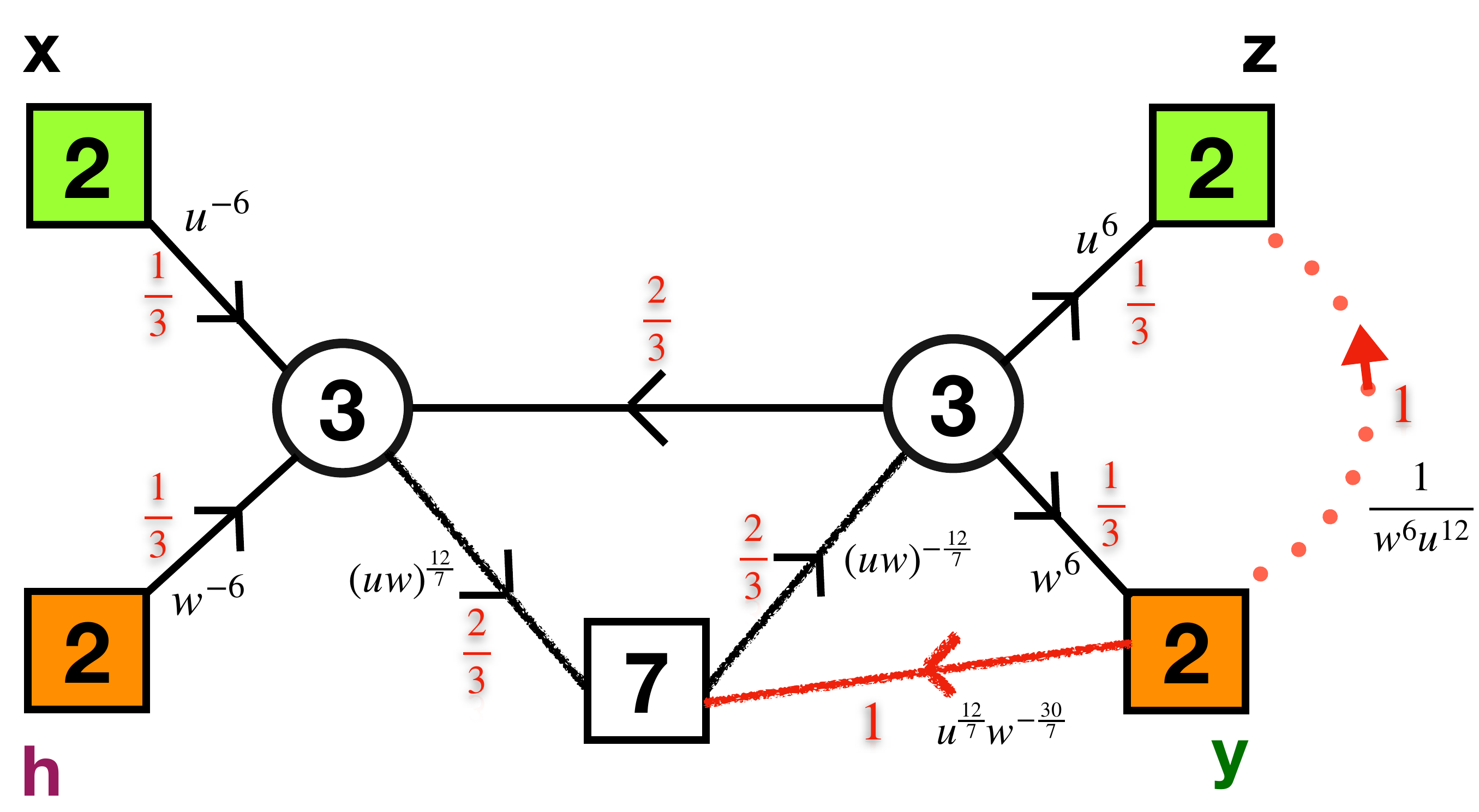}
    \caption{The basic pair of pants duality of S-gluing. The duality exchanges the $y$ and $h$ punctures. To get the punctures to be of the same type exactly we need to flip the moment maps of one of them. The flip fields are denoted in red. The dotted line flips the baryonic operator with charge $w^6u^{12}$. This field is in the fundamental representation of the symmetry it emanates from and a singlet of the symmetry it points to. The various superpotentials are detailed in the bulk of the paper.}
    \label{F:sdual}
\end{figure}

If we S-glue two copies of the same sphere together we obtain a surface with zero flux and four punctures of different type. Identifying the symmetries of the two spheres with conjugation of the punctures moment map operators in this gluing make the punctures different than one another.
To study dualities it is useful to have the same types of punctures which are exchanged by the various pairs of pants decompositions. We can however easily ``flip'' a puncture so that the moment maps will be conjugated by simply coupling the moment maps to an octet of chiral fields. The added superpotential term will set the original moment map to zero in the chiral ring \cite{Barnes:2004jj}  (see also \cite{Benvenuti:2017lle}), and the added octet of chiral fields will play the role of the new moment maps. \footnote{Such flips of punctures, or ``signs'' of punctures, are very common in various geometric constructions. See for example \cite{Agarwal:2015vla,Maruyoshi:2016tqk}.} See Figure \ref{F:sdual} for a quiver description of the duality. Note that the deformation with flip fields is relevant as the moment maps in the S-glued four puncture sphere have superconformal R-charge one. After the flip the superconformal R-charge changes. 

 Following the above procedure we will have two pairs of punctures of the same types, such that exchanging one pair corresponds to non equivalent pair of pants decomposition. We can test the duality by looking at the index \cite{Kinney:2005ej,Romelsberger:2005eg,Dolan:2008qi,Rastelli:2016tbz} which should be invariant under the exchange of fugacities of the same types of punctures. This is a non trivial mathematical fact as the computation itself is not obviously invariant. One can compute the index expansion in fugacities and show that it is indeed invariant under the above exchange. A related interesting question is to study the structure of the conformal manifold. From the index we can read off the marginal operators \cite{Beem:2012yn} to be in the following representations of the symmetry groups,
\be
{\bf 1}\;\oplus\; \left({\bf 2}_{y_1}\otimes {\bf 2}_{y_2}\otimes {\bf 2}_{z_1}\otimes {\bf 2}_{z_2}\right)\;\oplus  	\; {\bf 1}_{u^{12}/w^{12}}\;\oplus  	\; {\bf 1}_{w^{12}/u^{12}}\,,
\ee 
where each pair of fugacities $y_i$ or $z_i$ correspond to two punctures of the same type. Note that the superconformal R-symmetry is not the $6d$ one in this case as we added flipping fields to flip half the punctures. We see that there is a one dimensional direction on the conformal manifold on which all the symmetries are preserved and that is the natural one to be associated with the complex structure modulus. We have an additional direction on which the puncture symmetries are not broken but $w$ is locked to be the same as $v$. Finally we have exactly marginal directions on which the puncture symmetries are broken but these are not of interest to us.
 
To summarize: the four punctured sphere described above is a theory built from two copies of $SU(3)$ SQCD with $N_f=7$ connected by a single bifundamental field, with a particular superpotential,
and flip fields. This theory has an S-duality acting on its conformal manifold. We will show in the next section that in fact this duality is a generalization of the Intriligator-Pouliot duality of $SU(2)$ SQCD with eight fundamental fields \cite{Intriligator:1995ne}.
 
\subsubsection*{$\Phi$-gluing dualities}

$\Phi$-gluing two copies of the same trinion we obtain a sphere with one unit of flux and four punctures which are equal in pairs. We thus should expect that the model is invariant under the exchange of the same type of punctures. This model is very different from the one obtained in the S-gluing. The S-glued model can be obtained from this model by giving masses to the fundamental octet we introduced in the gluing. We can also as before check the duality property using the index. Computing the index we find that in expansion in fugacities it is invariant under the exchange 
of the same type punctures. Moreover, as $u,\,w$ and $v$ mix with the R-symmetry we find a single exactly marginal operator which is a singlet of the $SU(6)\times U(1)^3$ global symmetry. This is the exactly marginal operator which is natural to associate with the geometric modulus of the sphere. Here the model that enjoys the duality in the UV has the structure of $SU(3)^2\times SU(2)$ gauge theory such that the $SU(3)$ factors have six flavors, the $SU(2)$ has seven flavors and we have various superpotentials. 
As we have seen, although the $SU(2)$ gauging is naively irrelevant, turning on the various deformations sequentially, we find it is actually dangerously irrelevant. Using this deformation sequence we flow to an interacting SCFT in the IR which enjoys the duality discussed here.

Computing the index of a genus two surface using $\Phi$ gluing for all punctures we obtain,
\be
1+qp\,\left(3+\left\{1+{\bf 133}_{E_7}\right\}+3\frac1{u^{12}v^{12}w^{12}}+2\frac1{u^6v^6w^6}{\bf 56}_{E_7}\right)+\cdots\,.
\ee 
Note that the operators with negative $uvw$ charge are relevant (as can be seen from the mixing of \eqref{mixe7}), zero charge are marginal, and if we had operators with positive charge they would be irrelevant. The index has the expected structure  \cite{Razamat:2019ukg,Kim:2017toz}. 
In particular, on general grounds one expects 
that the index of a genus $g$ theory with flux ${\cal F}$ for a $U(1)$ breaking $E_8$ to $E_7\times U(1)$ to have the structure \cite{babuip,Beem:2012yn},
\be
&&1+qp\,\left(3g-3+(g-1)\left\{1+{\bf 133}_{E_7}\right\}+(g-1+2{\cal F})\frac1{u^{12}v^{12}w^{12}}\right.+\\
&&\;\;\;\;\;\left.(g-1-2{\cal F})u^{12}v^{12}w^{12}+(g-1+{\cal F})\frac1{u^6v^6w^6}{\bf 56}_{E_7}+(g-1-{\cal F})u^6v^6w^6{\bf 56}_{E_7}\right)+\cdots\,,\nonumber
\ee 
Plugging $g=2$ and ${\cal F}=1$ we reproduce the expression above modulo a missing factor of $-v^{12}u^{12}w^{12}$. This is a known peculiarity of the genus two E-string compactification \cite{Razamat:2019vfd}.
Another instance of a peculiarity for a genus two E-string compactification can be seen in \eqref{indexgen2}, where the flux is zero and we get an additional, ``accidental'', exactly marginal deformation.    

\subsubsection*{Closing a puncture}

Given a theory with a  puncture we can ``close'' it by giving a vacuum expectation value (vev) to a component of the moment map and adding certain fields and superpotential terms \cite{Kim:2017toz}. The resulting theory will correspond  to a surface with the puncture removed and the flux shifted. Let us assume that the charges of the octet of moment map operators are $u_i x^{\pm1}$ where the $u_i$ are combinations of fugacities for the Cartan of $E_8$ and $x$ is the Cartan of the puncture $SU(2)$. Let us denote the moment map component charged $u_ix^{\pm1}$ as $M^{\pm}_i$. Then giving a vev to $M_1^+$ for example we also introduce chiral superfields $F_i$ and couple them through the superpotential,
\be
W= M^-_1F_1+\sum_{i=2}^8 M^+_i F_i\,.
\ee 
To fully specify the flux of the system we need to specify the flux of all the Cartan generators of $E_8$. There are many choices of this octet of $U(1)$ symmetries. Given a puncture a natural choice is in terms of the eight $U(1)$s rotating each one of the components of the moment map. Thus the flux is specified by a vector ${\cal F}$ which has eight components.
These eight $U(1)$s are naturally the Cartans of $U(8)\subset SO(16)\subset E_8$.
Let us write the flux in terms of the $U(1)$s corresponding to the puncture we are closing. Then the flux shift after closing the puncture with vev to $M_1^+$, following the notations of \cite{Kim:2017toz} is,
$$\Delta {\cal F} =(2,0,0,0,0,0,0,0)\,.$$
 The shift of flux for each $U(1)$ is just proportional to the charge of the operator which received the vev under that $U(1)$. 
 
As we have seen from comparing anomalies the flux corresponding to the trinion is half a unit for a $U(1)$ breaking $E_8$ into $E_7\times U(1)$. We have three punctures and each has two baryonic moment maps and six mesonic ones. Let us write the vector of fluxes such that the first two components correspond to the baryonic operators and the last six to the mesonic ones.
Then we claim that the vector of fluxes for the trinion computed using the symmetries of either one of the three punctures is
 \be\label{trinflux1}{\cal F}=(-1,-1,0,0,0,0,0,0)\,.\ee
 This flux value can be motivated by matching anomalies of all the symmetries to the six dimensional predictions and by the consistency of the picture of flows we will describe below.
  This vector of fluxes preserves $E_7\times U(1)$ and with our normalization gives half a unit of flux for the $U(1)$. A way to read off the symmetry preserved by the flux is to find all the roots of $E_8$ which are orthogonal to it. See Appendix A of \cite{Pasquetti:2019hxf} for detailed explanation. For example here the roots of the $SO(16)$ subgroup are $(\pm1,\pm1,0,0,0,0,0,0)$ plus permutations. Out of these roots $\pm(1,-1,0,0,0,0,0,0)$ and $(0,0,\cdots)$ are orthogonal to ${\cal F}$. These roots build an $SU(2)\times SO(12)$ subgroup. To find the symmetry we also need to complete the $SO(16)$ to $E_8$ by adding the spinor
weights, which are $(\pm\frac12,\pm\frac12,\pm\frac12,\pm\frac12,\pm\frac12,\pm\frac12,\pm\frac12,\pm\frac12)$ with even number of minus signs. The ones which are orthogonal to ${\cal F}$ are of the form $\pm(\frac12,-\frac12,\cdots)$ with even number of minuses. These form the representation ${\bf 2}_{\SU(2)}\otimes {\bf 32}_{SO(12)}$. The spinorial weights complete the root system to the one of $E_7$. 

Let us now close the puncture with a baryonic moment map leading to the flux 
$${\cal F}+\Delta{\cal F} = (1,-1,0,0,0,0,0,0)\,.$$ This is again half a unit of $E_7\times U(1)$ flux related to the one we started with by an action of the Weyl group of $E_8$.
On the other hand if we give a vev to one of the mesonic moment maps, for example the first we obtain,
$${\cal F}+\Delta{\cal F} = (-1,-1,2,0,0,0,0,0)\,,$$   which is half a unit of flux breaking $E_8$ to $E_6\times SU(2)\times U(1)$.

 \begin{figure}[t]
	\centering
  	\includegraphics[scale=0.32]{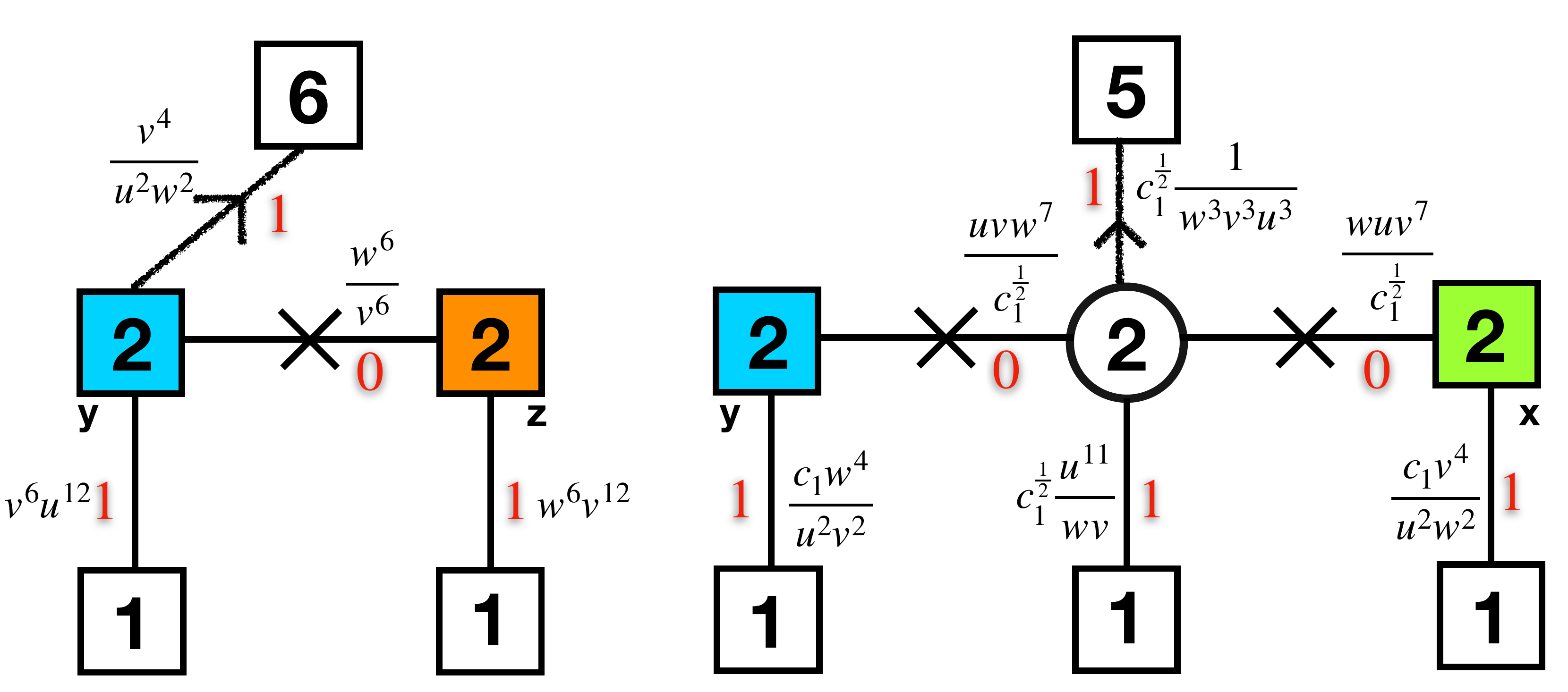}
    \caption{On the left we have the model obtained by closing a puncture with a baryonic vev. This is a tube with flux breaking $E_8$ to $E_7\times U(1)$. On the right we have the result of a mesonic vev. The theory corresponds to a tube with flux breaking $E_8$ to $E_6\times SU(2)\times U(1)$. The crosses are flip fields flipping the baryons built from the bi-fundamentals.}
    \label{F:tubes}
\end{figure}
We can now study explicitly the RG flows generated by giving various vevs for the trinion. A baryonic vev will Higgs the $SU(3)$ gauge symmetry completely leaving a WZ model, while the mesonic vev will Higgs the $SU(3)$ gauge symmetry down to $SU(2)$. The flows are easy to analyze and the result is depicted in Figure \ref{F:tubes}. We indeed find precisely models corresponding to tubes with the right amount of flux which were derived in \cite{Kim:2017toz}.\footnote{ Note that relative to  \cite{Kim:2017toz} some components of the moment maps are flipped but this difference is related to making a choice in defining the {\it color} of a puncture. In particular by $\Phi$-gluing the tubes into tori this freedom disappears and we land precisely on the tori theories of \cite{Kim:2017toz}.} Combining such models into tori we should obtain theories with $E_7\times U(1)$ and $E_6\times SU(2)\times U(1)$ symmetries respectively. This was indeed checked in \cite{Kim:2017toz}.\footnote{Let us mention that the torus with minimal $E_7$ flux is a bit special as, although the index exhibits the expected $E_7$ representations, the theory does not have a locus with this symmetry on the conformal manifold, as discussed in \cite{Kim:2017toz}.
In addition, the tori built from the $E_6\times SU(2)\times U(1)$ tube can be simplified by flows, producing for example the fact that $SU(2)$ SQCD with $N_f=4$ and particular gauge singlets has $E_6\times U(1)$ global symmetry \cite{Razamat:2017wsk}.}

We can perform another interesting exercise by flipping one of the components of the moment map octet and then closing it. Flipping does not change the flux but changes the type of puncture and in particular it conjugates the charges of the flipped component. Say we flip a mesonic component for starters, then the vector of fluxes is unchanged as ${\cal F}$ of \eqref{trinflux1} has zeros at the location of the mesonic components. Note that we choose our basis according to the puncture moment maps after the flipping of the mesonic component. After giving the vev the flux by the general rules above is shifted to,
$${\cal F}+\Delta{\cal F} = (-1,-1,2,0,0,0,0,0)\,,$$ which is again half a unit of flux breaking $E_8$ to $E_6\times SU(2)\times U(1)$. However, note that the field theory operation is rather different here as we give a vev to the flip field which simply gives a mass to one of the mesons resulting in $SU(3)$ $N_f=5$ SQCD (plus some gauge singlet fields). Without flipping we obtained $SU(2)$ SQCD with five flavors as can be seen in Figure \ref{F:tubes}. However the two theories are equivalent as they are Seiberg dual to each other. Thus we note that the consistency of the picture hinges on the validity of Seiberg duality.

\begin{figure}[t]
	\centering
  	\includegraphics[scale=0.32]{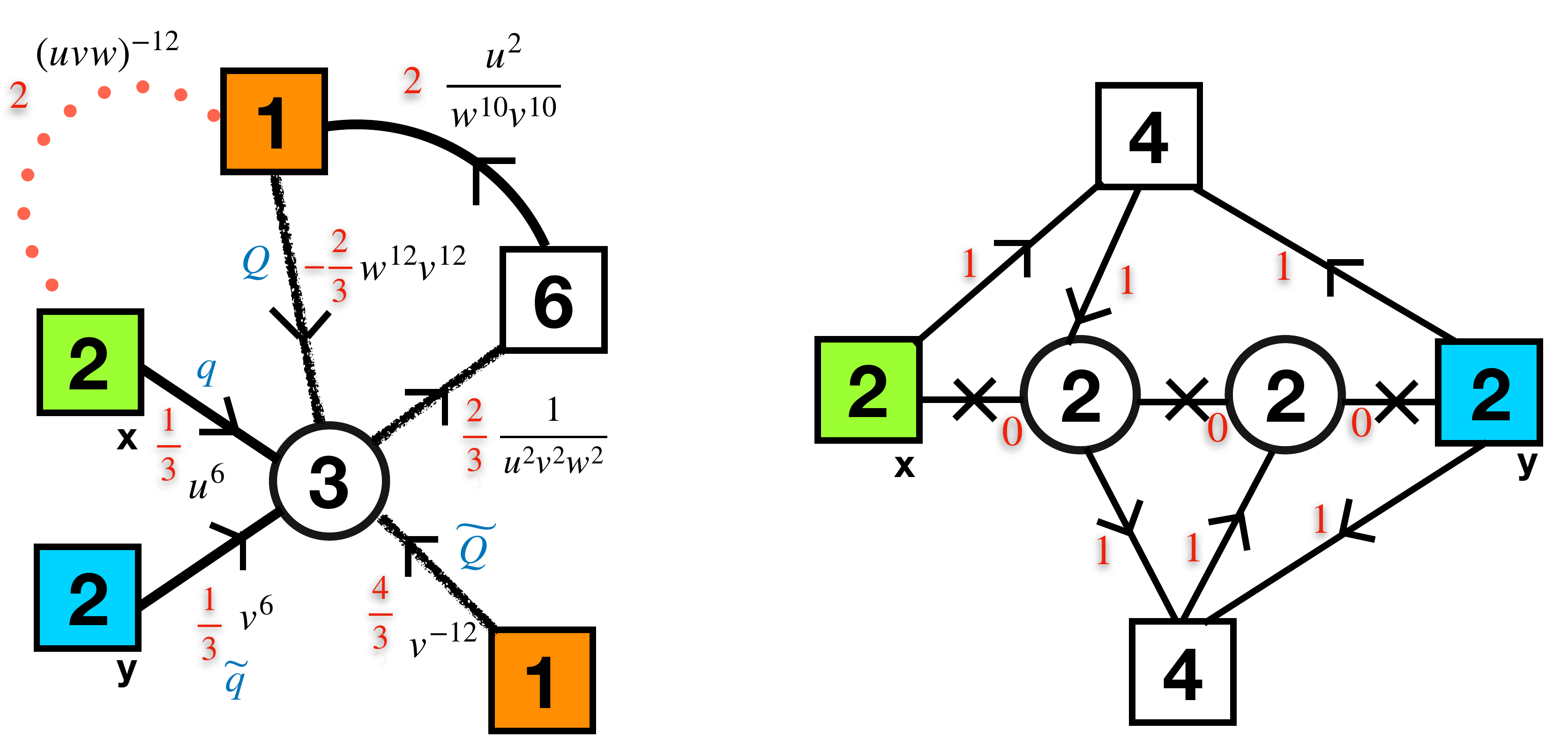}
    \caption{The tube theories for $SO(12)\times U(1)\times U(1)$ preserving minimal flux. On the right we have the tube derived in \cite{Kim:2017toz}
    by combining together three basic $E_7\times U(1)$ preserving tubes. On the left is the tube derived here by flipping one of the baryonic moment map components and giving it a vev.
    The dotted line corresponds to a chiral field flipping the baryonic operator of the form $q^2Q$. The theory has also a baryonic superpotential ${\widetilde q}^2\widetilde Q$. $\Phi$-gluing two copies of each one of the tubes into a torus by gauging two $SU(2)$ puncture symmetries and flipping the moment maps one obtains a pair of dual theories. This duality is a consequence of the geometric construction.
    }
    \label{F:baryonflip}
\end{figure}

Finally we can flip a baryonic moment map and give a vev to the flip field. Here something more interesting happens. Giving vev to the flip field generates a superpotential term which is the baryon. Thus, the theory is $SU(3)$ SQCD with $N_f=6$, and a baryonic superpotential for one of the baryons (plus gauge singlet fields). Note that as we flipped the baryons the vector of fluxes written in terms of the symmetries of the new puncture is $(1,-1,0,0,0,0,0,0)$.
Computing the vector of fluxes after closing the puncture we get,
\be
{\cal F}+\Delta{\cal F} = (3,-1,0,0,0,0,0,0)\,,
\ee 
which corresponds to the preserved symmetry $SO(12)\times U(1)\times U(1)$ with half a unit of flux in {\it both} $U(1)$s.
Two punctured spheres with such a flux were studied in \cite{Kim:2017toz}. In particular one can build them by combining three basic tubes with partial S and $\Phi$ gluing.
The tube that was obtained in \cite{Kim:2017toz} has $SU(2)^2$ gauge group and is depicted in Figure \ref{F:baryonflip}. The tube we get here and the one from \cite{Kim:2017toz} might differ by the choice of types of punctures, however once we glue say pairs of such tubes in to a torus this difference is inessential and we get a pair of dual theories: one has $SU(2)^6$ gauge group and another $SU3)^3\times SU(2)^2$ and a baryonic superpotential. One can easily check that the conformal anomalies and indices indeed agree.\footnote{We have explicitly matched the index in expansion in the superconformal fugacities and the $U(1)$ which mixes with the $6d$ R-symmetry.} This duality is needed for consistency of the picture and indeed it seems to be correct.

We also note the following curious fact.  The trinion is the $SU(3)$ SQCD  with $N_f=6$. We have a two punctured sphere which has $N_f=5$ and we can further flip and close additional puncture to get a one punctured sphere with $N_f=4$. Moreover, we also have seen that spheres with four punctures and say zero flux have $SU(3)$ gauge theory ingredients with $N_f=7$, the five puncture (and higher) spheres will have $SU(3)$ ingredients with $N_f=8$. Models with higher genus  having trinions with the three puncture symmetry gauged will have $SU(3)$ gauge nodes with $N_f=9$ but no higher. Thus we cover all the possible numbers of flavors asymptotically free $SU(3)$ gauge theory can have in these geometric constructions. 

\section{Minimal D-conformal matter trinion and compactifications}\label{sec:D}

The E-string $6d$ SCFT can be thought of as the theory residing on a single M5 brane probing a $D_4$ singularity. As such it is the first model in a sequence of interacting SCFTs residing on a single $M5$ brane probing a $D_{N+3}$ singularity with $N\geq 1$.\footnote{The case of $N=0$ is special as it can be related to one M5 brane on a ${\mathbb Z}_4$ singularity, which is a free theory.} These models are called $(D_{N+3},D_{N+3})$ minimal conformal matter SCFTs \cite{DelZotto:2014hpa}. We will call these models minimal $D_{N+3}$ conformal matter for brevity. 
The theories residing on more than one brane are called non-minimal but we will not consider them here. Our construction for the E-string, minimal $D$ conformal matter with $N=1$, has a direct generalization for $N>1$. Let us mention that the global symmetry of these $6d$ SCFTs is $SO(4N+12)$ which enhances to $E_8$ for $N=1$.

\subsubsection*{The trinion}

The basic claim is that the ${\cal N}=1$ $SU(N+2)$ $N_f=2N+4$ SQCD is a trinion of minimal $D_{N+3}$ conformal matter with two maximal $SU(N+1)$ punctures and one minimal $SU(2)$ puncture.\footnote{The SQCD with special unitary groups in the middle of the conformal window has yet another geometric interpretation as a compactification
of $(2,0)$ theory on a sphere with two maximal and two minimal punctures with zero flux (in our notations) \cite{Benini:2009mz,Bah:2012dg}. In these compactifications only an $SU(N+2)^2\times U(1)^3$ subgroup of the full symmetry group has a geometric meaning as the symmetry associated to the punctures and to the Cartan of the extended R-symmetry of the $(2,0)$ theory.}  Note that the minimal $D_{N+3}$ conformal matter has in fact three different types of maximal punctures \cite{Kim:2018bpg}. This is related directly to the fact that the $5d$ reduction of this model with different holonomies has three different effective gauge theory descriptions \cite{Hayashi:2015fsa,Hayashi:2015zka}. The three types of puncture have rank $N$ symmetry which is either $SU(N+1)$, $USp(2N)$, or $SU(2)^N$ (for $N=1$ are all  the same). In \cite{Razamat:2019ukg} a trinion with two maximal $SU(2)^N$ and a minimal $SU(2)$ puncture was found and here we trade $SU(2)^N$ for $SU(N+1)$.
The $SU(2)$ minimal puncture in the trinion discussed here is the same as the one in \cite{Razamat:2019ukg}, and it can be obtained by partially closing a $USp(2N)$ maximal puncture, see \cite{Razamat:2019ukg} for details.

\begin{figure}[t]
	\centering
  	\includegraphics[scale=0.38]{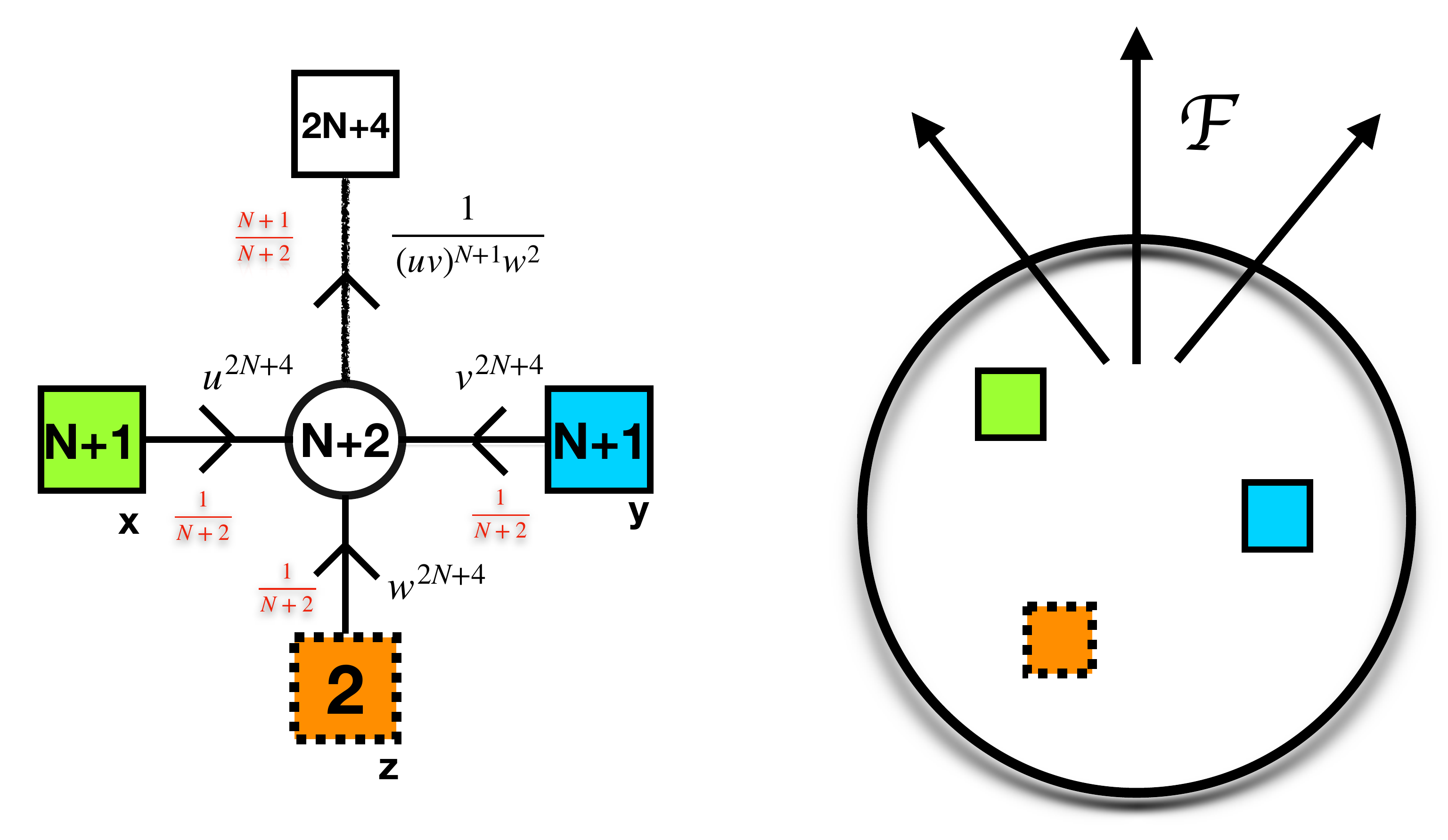}
    \caption{A trinion of the minimal D conformal matter. The choice of R-charges is denoted in red. These are the R-charges inherited from six dimensions and not the superconformal ones, which are a half for all the fields. There are no superpotential terms.}
    \label{F:trinionD}
\end{figure}

The SQCD model has a $U(1)$ baryonic symmetry and two copies of $SU(2N+4)$ symmetries rotating the fundamentals and the anti-fundamentals. We split the $SU(2N+4)$ rotating the fundamentals into $SU(N+1)\times SU(N+1)\times SU(2)\times U(1)^2$ and associate the three non-abelian factors to the three punctures. The remaining $SU(2N+4)\times U(1)^3$ symmetry is a sub-group of the $6d$ $SO(4N+12)$ global symmetry. We assign to the various fields non-anomalous R-charges as depicted in Figure \ref{F:trinionD}. This is the R-symmetry inherited from $6d$ and is not in general the superconformal one.
 
 As in the case of E-string each puncture has an interesting set of moment map operators in the fundamental representation of the puncture symmetry group.
 For both, minimal and maximal, punctures here these form a $(2N+6)$-plet of  fields.
The moment maps for the maximal punctures are,
\be\label{momentmapsN}
&&M_u={\bf N+1}^x\;\otimes\; \left({\bf 2N+4}_{u^{N+3}v^{-(N+1)}w^{-2}}\oplus {\bf 1}_{(u v^{N+1})^{2N+4}}\right)\;\oplus\;
\overline{{\bf N+1}}^x\;\otimes {\bf 1}_{(u^Nw^2)^{2N+4}}\,,\nonumber\\
&&M_v={\bf N+1}^y\;\otimes\; \left({\bf 2N+4}_{v^{N+3}u^{-(N+1)}w^{-2}}\oplus {\bf 1}_{(v u^{N+1})^{2N+4}}\right)\;\oplus\;
\overline{{\bf N+1}}^y\;\otimes {\bf 1}_{(v^Nw^2)^{2N+4}}\,.\nonumber\\
\ee
The moment map for the minimal puncture is,
\be\label{minmom}
&&M_w={\bf 2}^z\;\otimes\; \left({\bf 2N+4}_{(uv)^{-(N+1)}w^{2N+2}}\oplus {\bf 1}_{(w v^{N+1})^{2N+4}}\oplus {\bf 1}_{(w u^{N+1})^{2N+4}}\right)\,.
\ee
As in the case of the E-string, $2N+4$ of the moment maps are mesons and two are baryons. The nature of the operators can be easily inferred from their charges. 
Let us compute some of the anomalies associated to the trinion, 
\be
&&\Tr\, R= -N^2-4N-5\,,\;\; \;\;\;\; \Tr\, R^3= -N^2+2N+1\,,\;\;\\
&&  \Tr\, R\, SU(N+1)^2=\Tr \, R \,SU(2)^2 =-\frac12(N+1)\,.\nonumber
\ee 
Note that the puncture anomalies are as expected from $6d$ \cite{Kim:2018bpg}.

\subsection{Gluings}

\subsubsection*{S-gluing}
Next we consider combining the trinions. First we consider the S-gluing \cite{Kim:2017toz,Kim:2018bpg,Razamat:2019ukg}.
This is the procedure of gluing two punctures together by gauging the puncture $SU(N+1)$ symmetry and coupling the moment maps of the two punctures, $M$ and $M'$, through a superpotential,
\be\label{Esuperpotential}
W= \sum_{i=1}^{2N+6}  M_i M'_i\,.
\ee 
In particular if we take two identical trinions and S-glue them together the symmetries of one are identified with the conjugation of the other. This  implies that the flux of the combined surface is zero. Note that our R-charge is not anomalous under this gluing as the $SU(N+1)$ gaugings here have $N+2$ flavors.  As some of the moment maps are  mesons and some are  baryons, the superpotential has quartic and higher power ($2N+4$th power) terms. 

 \begin{figure}[htbp]
	\centering
  	\includegraphics[scale=0.38]{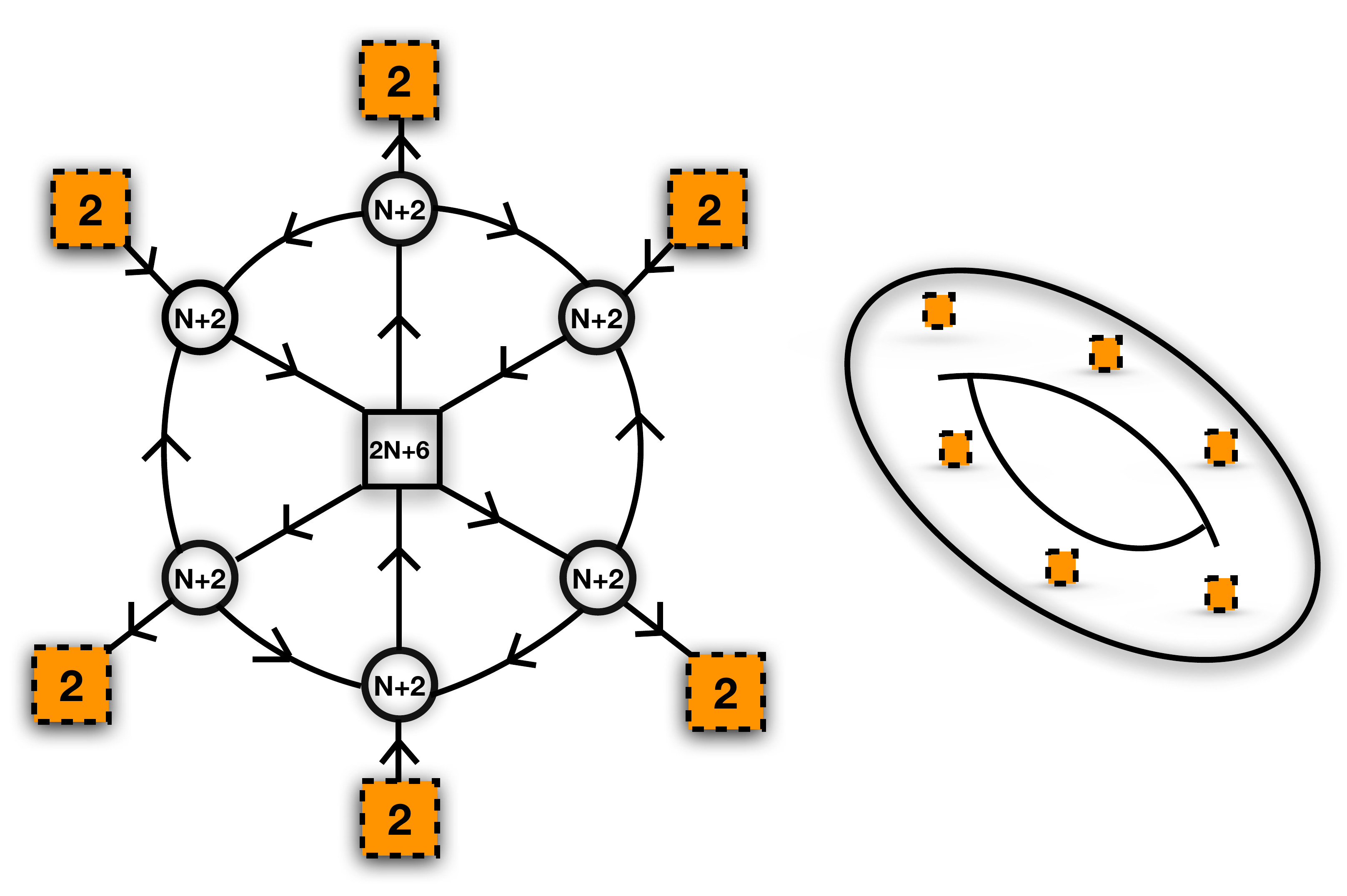}
    \caption{The minimal D-conformal matter SCFT compactified on a torus with an even number of minimal punctures and zero flux. The theory is obtained by S-gluing together an even number of trinions. We have baryonic superpotentials for the fields in the bi-fundamental representation of two gauge groups, and cubic superpotentials for the triangles of the ``wheel''. The case of $N=1$ is special as also baryonic superpotentials for the spikes of the ``wheel'' can be turned on.
    Taking $2N$ minimal punctures we will obtain a torus with two maximal $USp(2N)$ punctures \cite{Razamat:2019ukg} from which genus $g$ models can be constructed.}
    \label{F:wheelD}
\end{figure}
Note that as the $SU(N+1)$ gauging has $N+2$ flavors we can use Seiberg duality \cite{Seiberg:1994pq}. The dual theory is the WZ model of mesons and baryons. The former will give a bi-fundamental of the two $SU(N+2)$ gauge groups of the trinions and the latter an (anti)fundamental chiral. After one such gluing each $SU(N+2)$ gauge group will have $2N+5$ flavors. Combining several trinions into a circle one obtains the theory depicted in Figure \ref{F:wheelD}. This model is associated to a torus with some number of minimal punctures. As each trinion is glued to two others, each $SU(N+2)$ gauge group has now $2N+6$ flavors. 

Next we can utilize the observation of \cite{Razamat:2019ukg} that $N$ minimal $SU(2)$ punctures on some locus of the conformal manifold can combine into a $USp(2N)$ maximal puncture. Such recombinations of punctures were observed also in twisted compactifications of class ${\cal S}$ where they are referred to as {\it atypical degenerations} \cite{Chacaltana:2013oka,Chacaltana:2016shw,Chacaltana:2012ch}, see also \cite{Razamat:2016dpl,Razamat:2018gro}. As the observation  of \cite{Razamat:2019ukg} is a ``local'' statement about the properties of the punctures it should hold also in our case. We can construct a torus with $2N$ minimal punctures, combine these into two maximal $USp(2N)$ ones, and then S-glue the $USp(2N)$ punctures to form higher genus surfaces. In particular let us compute the conformal anomalies here. To form a genus $g$ surface we need to glue $2N$ trinions through the $SU(N+1)$ punctures into a torus with two maximal $USp(2N)$ punctures and then glue $g-1$ copies of such tori by gauging the $USp(2N)$ symmetries. The conformal anomalies of such a model are then,
\be
a= \frac3{16} N(16N+9)(g-1)\,,\qquad c= \frac18N(25N+18)(g-1)\,.
\ee 
These agree perfectly with the predicted anomalies from compactifying minimal $D_{N+3}$ conformal matter on a genus $g$ surface with zero flux \cite{Razamat:2019ukg,Kim:2018bpg}. 
This also motivates our R-charge assignment. Note that as we S-glue the trinions all the symmetries appear in conjugated pairs and thus do not mix with the R-symmetry. Thus, assuming the construction gives an SCFT with no accidental abelian symmetries the above R-symmetry indeed gives the conformal anomalies.\footnote{
The analysis of dynamics here is very similar to the $N=1$ case so we will not repeat it.}
In particular all the symmetries should enhance for genus $g$ compactifications with zero flux to $SO(4N+12)$ and then the only R-symmetry is the one inherited from six dimensions.

\subsubsection*{$\Phi$-gluing}
Next we consider $\Phi$-gluing \cite{Kim:2017toz,Kim:2018bpg,Razamat:2019ukg}. This is done by gauging the puncture symmetry and identifying the moment map operators of the two punctures using the superpotential,
\be
W=\sum_{i=1}^{2N+6} \left(M_i-M_i'\right)\Phi_i\,,
\ee 
where $\Phi_i$ is a collection of fields in the fundamental representation of the gauged symmetry which are added to the model. Note here as the symmetries of the glued punctures are identified without complex conjugation the flux of the glued surfaces is summed up. In particular this allows us to fix the flux of the trinion by comparing to anomalies predicted from six dimensions.
The analysis of the dynamics is naively similar to the $N=1$ case however is more complicated and we will not perform it in detail here.\footnote{For example for general value of $N$ some of the operators might hit  unitarity bounds and decouple \cite{Kutasov:2003iy}. In what follows we ignore such important effects  and leave the analysis of the dynamics for general $N$ with general value of flux for future work. Thus, as discussed in footnote 5,  the quoted statements of $a$ and $c$ anomalies should be taken as matchings of certain `t Hooft anomalies.   Note that the $SU(N+1)$ gauging in the UV free fixed point is IR free for $N=1$, marginal for $N=2$, and asymptotically free for $N>2$. At the fixed point of the trinions, where the chiral fields of the trinion have R-charges $+\frac12$, the gauging is IR free for $N<6$, marginal for $N=6$, and  asymptotically free for $N>6$.}  We  will  assume that the $\Phi$-glued theories make sense as  SCFTs to compute the trial $a$ and $c$ anomalies of a genus $g$ surface. 

The anomaly polynomial computation of a torus with $2N$ minimal punctures is straightforward: one computes the anomalies of $2N$ trinions, adds $N$ contributions for each of the two types of flip fields for the two types of puncture moment maps, and the contribution of $2N$ $SU(N+1)$ gauge fields. With this anomaly we now treat $N$ minimal punctures as a $USp(2N)$ maximal one
and thus the anomaly we computed is also the one for a torus with two maximal $USp(2N)$ punctures. 
We $\Phi$-glue $g-1$ such tori by gauging $g-1$ $USp(2N)$  symmetries and adding flip fields for the matching moment maps. The moment maps for the $USp(2N)$ maximal puncture are in the fundamental representation of $USp(2N)$, there are $2N+6$ of them, and their charges under other $6d$ symmetries are just the charges of the $SU(2)$ minimal puncture \eqref{minmom} \cite{Razamat:2019ukg}. All in all performing a-maximization of the result we obtain the following expression,
\be
&&a=\frac{4 N\left(10 N^2+22 N+13\right)^{3/2}+9 \left(16 N^3+53 N^2+56 N+16\right) N}{48 (N+2)^2}\,,\\
&&c=\frac{2 \left(11 N^2+26 N+17\right) \sqrt{N^2 \left(10 N^2+22 N+13\right)}+3 \left(25 N^3+86 N^2+98 N+34\right) N}{24 (N+2)^2}\,. \nonumber
\ee
This agrees with the anomalies of the minimal $D_{N+3}$ conformal matter compactified on genus $g$ surface with $N(g-1)$ units of flux in the $U(1)$ breaking the $SO(4N+12)$ symmetry to 
$SO(4N+8)\times SU(2)\times U(1)$ computed in \cite{Razamat:2019ukg,Kim:2018bpg}.\footnote{In notations of \cite{Razamat:2019ukg,Kim:2018bpg} this is the $r=2$ case of equations (2.6) and (2.7) in \cite{Razamat:2019ukg}.} As we have glued together $2N(g-1)$ trinions this implies that the flux of a single trinion is $\frac12$ for the above $U(1)$. Note that in the $N=1$ case the $SO(12)\times SU(2)$ symmetry enhances to $E_7$.  In principle one can now start gluing various flux
tubes of \cite{Kim:2018bpg} and check the anomalies and symmetry enhancements in index computations. For general flux the expressions become quite cumbersome so we will refrain from doing so.
 
The $U(1)$ with the above flux is a combination of $u$, $v$, and $w$ as in the E-string case. The specific combination is $\left((vu)^{N+1}w^2\right)^{N+2}$, where we see that it reproduces the E-string result for $N=1$.
The R-symmetry obtained using a-maximization is,
\be
R^{g=2}= R+\frac{\sqrt{10 N^2+22 N+13}-3}{6 (N+2)^3}\left(q_u+q_v+q_w\right)\,.
\ee
The branching rules are given by
\be
SO(4N+12) & \to & SO(4N+8) \times SU(2)_{\left(\frac{v}{u}\right)^{(N+1)(N+2)}} \times SU(2)_{\left((vu)^{N+1}w^2\right)^{N+2}} \nonumber\\
 & \to & SU(2N+4) \times U(1)_{\left(\frac{w^2}{uv}\right)^{N+1}} \times SU(2)_{\left(\frac{v}{u}\right)^{(N+1)(N+2)}} \times SU(2)_{\left((vu)^{N+1}w^2\right)^{N+2}},\nonumber\\
\textbf{Adj}_{SO(4N+12)} & \to & \textbf{Adj}_{SO(4N+8)} \oplus \textbf{3}_{SU(2)_1} \oplus \textbf{3}_{SU(2)_2} \oplus \left(\textbf{2}_{SU(2)_1}\otimes \textbf{2}_{SU(2)_2} \otimes \textbf{V}_{SO(4N+8)}\right), \nonumber\\
\textbf{Adj}_{SO(4N+8)} & \to & \textbf{Adj}_{SU(2N+4)} \oplus \textbf{AS}_{SU(2N+4)}\left(\frac{w^2}{uv}\right)^{2(N+1)} \oplus \overline{\textbf{AS}}_{SU(2N+4)}\left(\frac{uv}{w^2}\right)^{2(N+1)} \oplus \textbf{1}, \nonumber\\
\textbf{V}_{SO(4N+8)} & \to & \textbf{(2N+4)}_{SU(2N+4)}\left(\frac{w^2}{uv}\right)^{N+1} \oplus \overline{\textbf{2N+4}}_{SU(2N+4)}\left(\frac{uv}{w^2}\right)^{N+1}\,,
\ee
where $\textbf{Adj},\,\textbf{AS}$ and $\textbf{V}$ denote the adjoint, two index antisymmetric and vector representations, respectively.
As an example we consider the genus two theory of vanishing flux obtained by $S$-gluing $2N$ trinions. Calculating its supersymmetric index to leading order we find for $N>1$,
\be
1 + \left(3+\textbf{Adj}_{SO(4N+12)}\right)pq+...,
\ee
with the $\textbf{Adj}_{SO(4N+12)}$ decomposed as shown above into the UV symmetries. As expected we see the emergence of the full $6d$ $SO(4N+12)$ symmetry in the IR.
 
\subsection{Dualities and closing punctures}

The discussion here is very similar to the one for the E-string on one hand, but explicit checks are rather cumbersome on the other hand. Thus, for the sake of brevity the discussion will be terse.

\subsubsection*{Dualities}

We have understood how to build the trinions and how to glue them. The resulting surfaces should correspond to Riemann surfaces and thus as long as we obtain the same surface the ingredients and their composition should not matter, leading to many dual descriptions. One of the basic dualities one can construct is, as for the E-string, to take two trinions and $S$-glue them together flipping one of the unglued maximal (or minimal) punctures. The exchange of the two punctures is a different pair of pants decompoistion of the surface and thus constitutes a duality. For example, the index has to be invariant under it. Similarly we can $\Phi$ glue theories together and should find dual descriptions when exchanging punctures of the same type. Moreover in \cite{Kim:2018bpg} two punctured spheres with various values of flux for the minimal $D_{N+3}$ conformal matter were derived and we can admix them in various ways as well as the trinion with two $SU(2)^N$ and one $USp(2N)$ puncture found in \cite{Razamat:2019ukg}.  For example we can build a torus with two $USp(2N)$ punctures from the latter trinions and glue this to the theories we obtain in this paper. Again as long as the topology, the flux, and the types of punctures are the same the models should be dual to each other.

 \begin{figure}[htbp]
	\centering
	\includegraphics[scale=0.23]{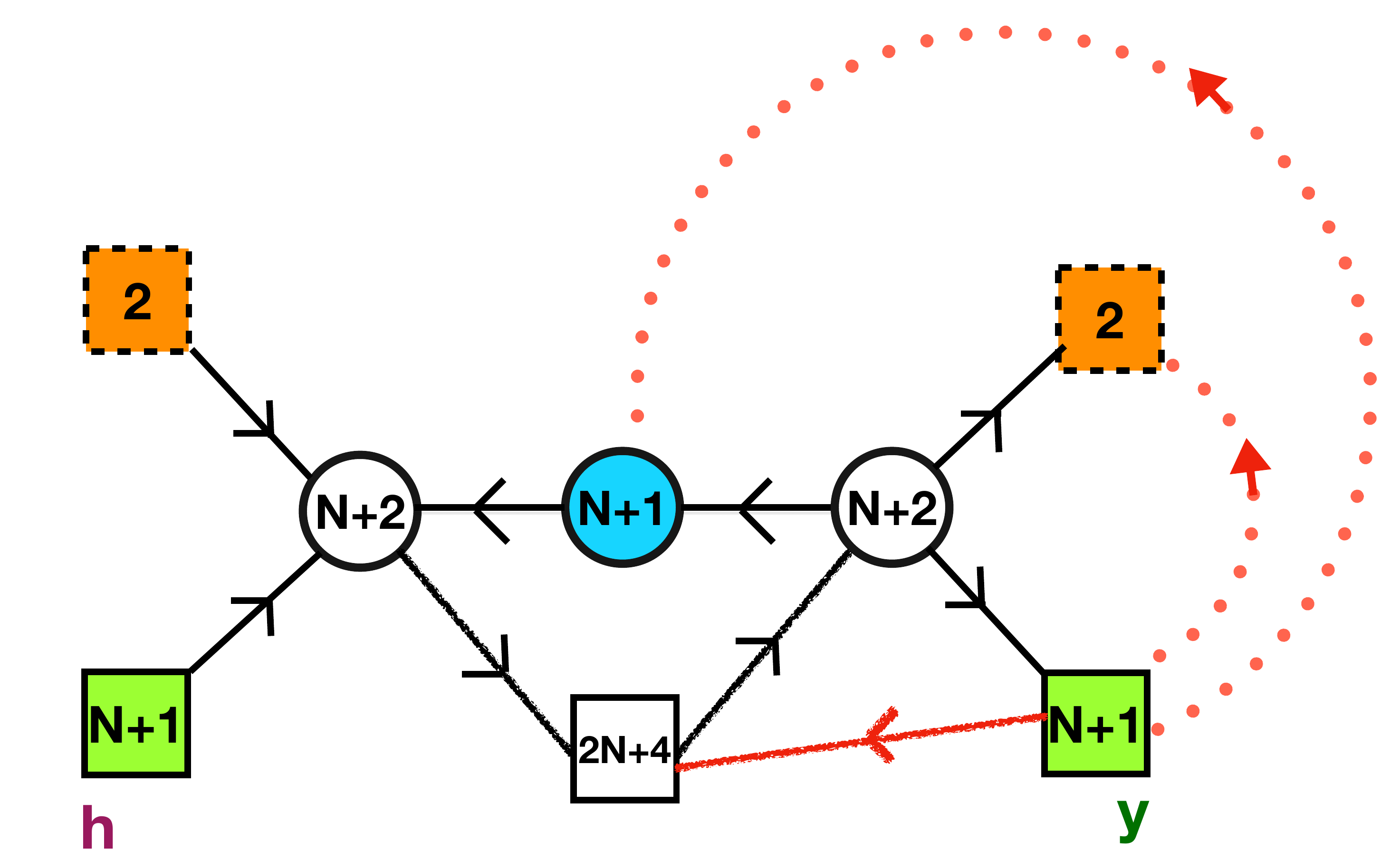}\;\;\; \includegraphics[scale=0.23]{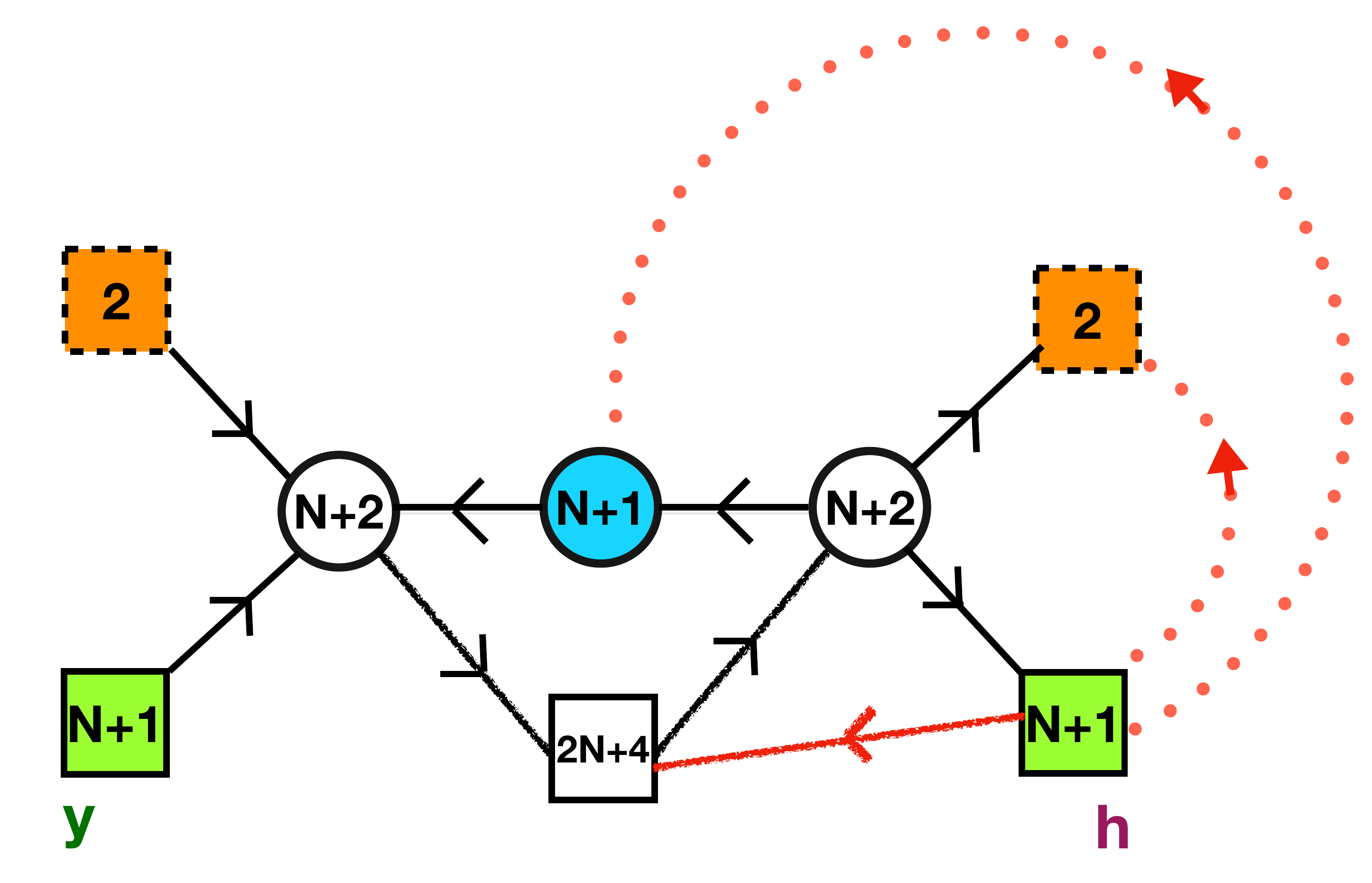}\\
  	\includegraphics[scale=0.23]{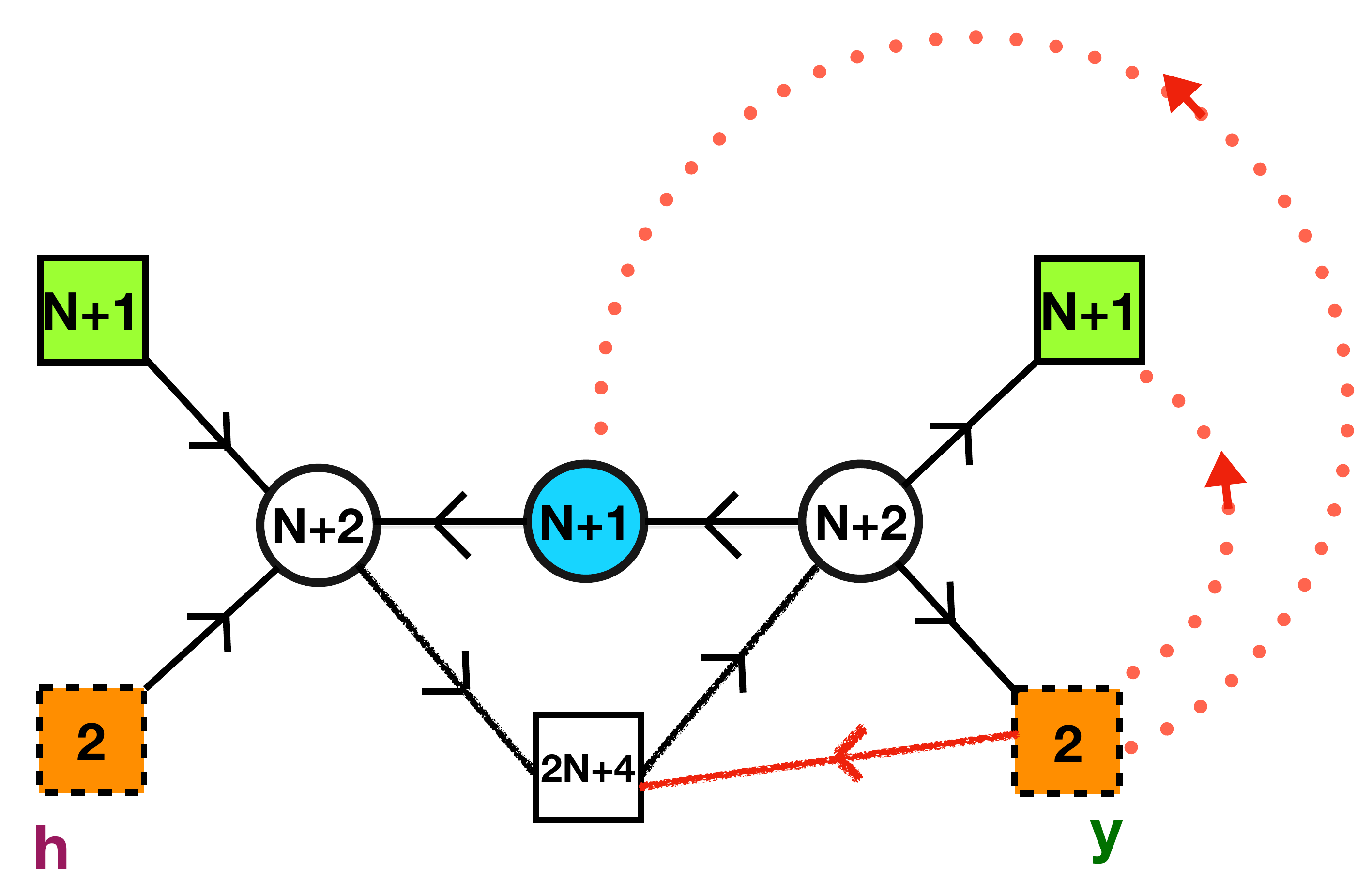}\;\;\; \includegraphics[scale=0.23]{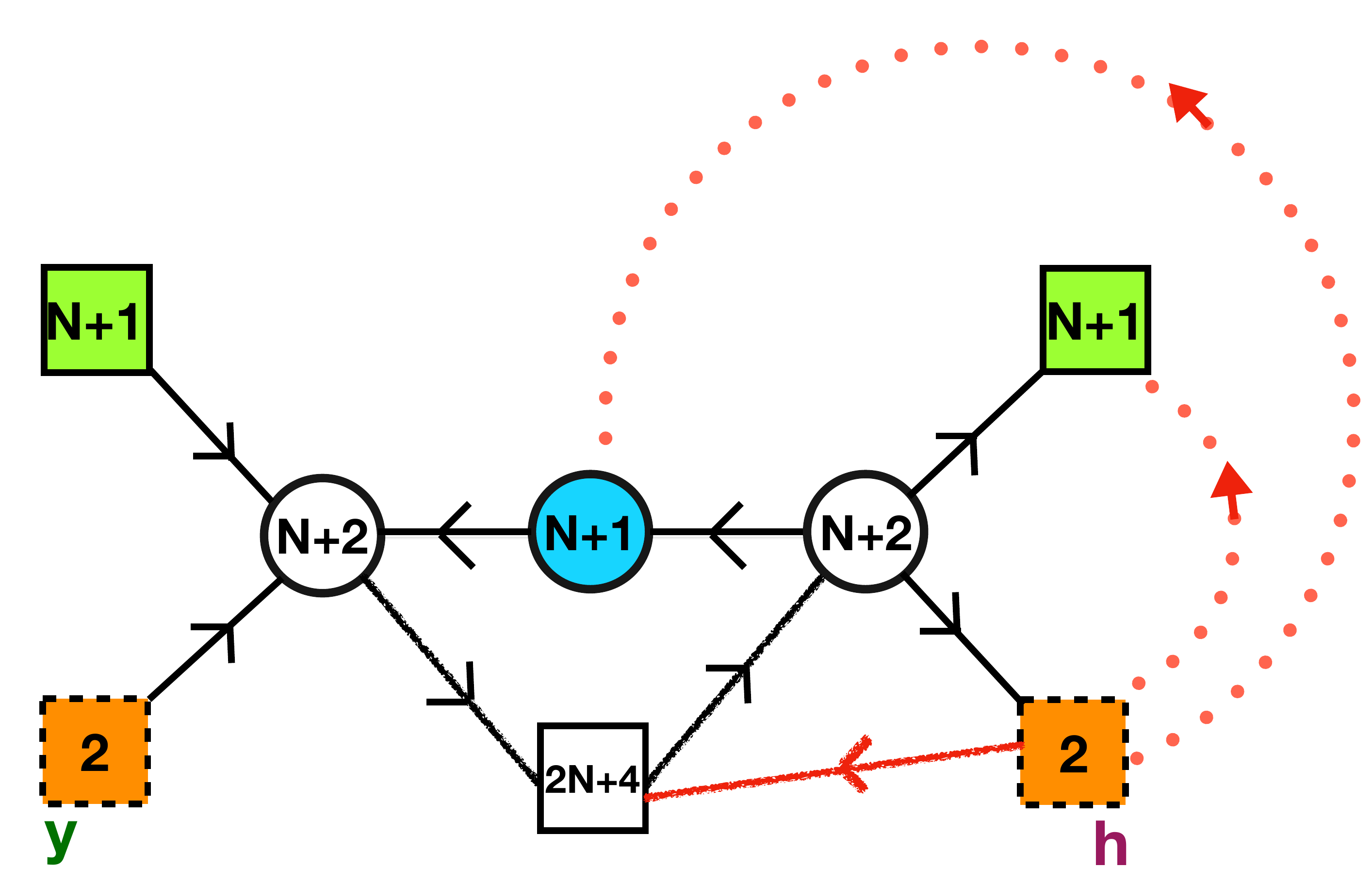}
    \caption{The basic pair of pants dualities of S-gluing. The duality exchanges the $y$ and the $h$ punctures. On the top we have the exchange of two maximal punctures and on the bottom the minimal punctures. For the punctures to be of the same type exactly we need to flip the moment maps of one of them. The flip fields are denoted in red. The dotted lines flip the baryonic operators. This field is in the fundamental representation of the symmetry it emanates from and a singlet of the symmetry it points to.
    In particular the dashed lines in the Figure are singlets of the gauge symmetry group. The various superpotentials are detailed in the bulk of the paper. Note that usually we can only flip maximal punctures but here as collections of minimal punctures build a maximal puncture we can do the same for minimal ones.}
    \label{F:IPdual}
\end{figure}

Let us make the following observation. Our discussion now has $N$ as a parameter which in principle can be taken to zero. The case of $N=0$ is related to one $M5$ brane probing a $D_3\sim A_3$ singularity which is a free SCFT. Thus this case is rather degenerate. However, let us proceed with our claims also to $N=0$. Here the maximal puncture symmetry is formally $SU(1)$ and thus we do not have gauging when gluing punctures. The minimal puncture becomes ``bigger'' than the maximal one and has an $SU(2)$ symmetry.
 Then the theory obtained by S-gluing two trinions depicted in Figure \ref{F:IPdual} becomes  just two copies of $SU(2)$ SQCD with eight fundamentals glued together with a superpotential.\footnote{This is reminiscent of the construction of \cite{Dimofte:2012pd} where it was shown that this model has an $E_7$ symmetry at a point on its conformal manifold. In \cite{Razamat:2017hda} it was argued that actually at that point a new $U(1)$ emerges and the symmetry is $E_7\times U(1)$.} We claim that the duality that exchanges the minimal punctures holds true also in this case. This follows directly by applying the Intriligator-Pouliot duality  \cite{Intriligator:1995ne} on both $SU(2)$ gauge nodes. This duality, in contrast to the usual Seiberg duality, flips both the mesons and the baryons. It is easy to see that with the flip fields already in place the duality will move them to the other node. Thus, we can view the Intriligator-Pouliot duality as the geometric statement of exchanging the two minimal punctures in this degenerate case. The non-degenerate higher $N$ cases can be then viewed as higher rank generalizations of Intriligator-Pouliot duality.

\subsubsection*{Closing a puncture}

As for the E-string, or any other compactification (see {\it e.g.} \cite{Gaiotto:2015usa,Bah:2017gph}), given a theory with a minimal $SU(2)$ puncture we can ``close'' it by giving a vacuum expectation value (vev) to a component of its moment map and adding certain fields and superpotential. Note that here we have more than one type of puncture, minimal is not the same as maximal, and there is a variety of punctures in between. We will first focus on closing minimal to no puncture and comment on closing maximal to smaller punctures in the end.
Let us assume that the charges of the $2N+6$-plet of moment map operators are $u_i x^{\pm1}$, where $u_i$ are combinations of fugacities for the Cartan of $SO(4N+12)$ and $x$ is the Cartan of the minimal puncture $SU(2)$. Let us denote the moment map component charged $u_ix^{\pm1}$ as $M^{\pm}_i$. Then giving vev to say $M_1^+$ we also introduce chiral superfields $F_i$ and couple them through the superpotential,
\be
W= M^-_1F_1+\sum_{i=2}^{2N+6} M^+_i F_i\,.
\ee 
To fully specify the flux of the system we need to specify the flux of all of the Cartan generators of $SO(4N+12)$. Here it is given by a vector ${\cal F}$ which has $2N+6$ components.
These $2N+6$ $U(1)$s are naturally the Cartans of $U(2N+6)\subset SO(4N+12)$.
Let us write the flux in terms of the $U(1)$s corresponding to the puncture we are closing. Then the shift in flux after closing the puncture is, following the notations of \cite{Kim:2017toz},
$$\Delta {\cal F} =(2,0,0,0,0,\cdots)\,.$$
 The shift of flux for each $U(1)$ is just proportional to the charge of the operator which received the vev under that $U(1)$. 
 
As we have seen from comparing anomalies the flux corresponding to the trinion is half a unit in $U(1)$ breaking $SO(4N+12)$ into $SO(4N+8)\times SU(2)\times U(1)$. We have one minimal puncture with two baryonic moment maps and the rest are mesonic. Let us write the vector of fluxes such that the two first components correspond to the baryonic operators and the last $2N+4$ to the mesonic ones.
Then we claim that the vector of fluxes computed using the symmetries of either one of the three punctures is
 \be\label{trinflux}{\cal F}=(-1,-1,0,0,0,0,0,\cdots)\,.\ee Since the $6d$ symmetry is not enhanced as it is for $N=1$, we only need to find the roots of $SO(4N+12)$ orthogonal to the vector of fluxes to determine the preserved symmetry and it is indeed $SO(4N+8)\times SU(2)\times U(1)$.
 
  \begin{figure}[htbp]
	\centering
  	\includegraphics[scale=0.32]{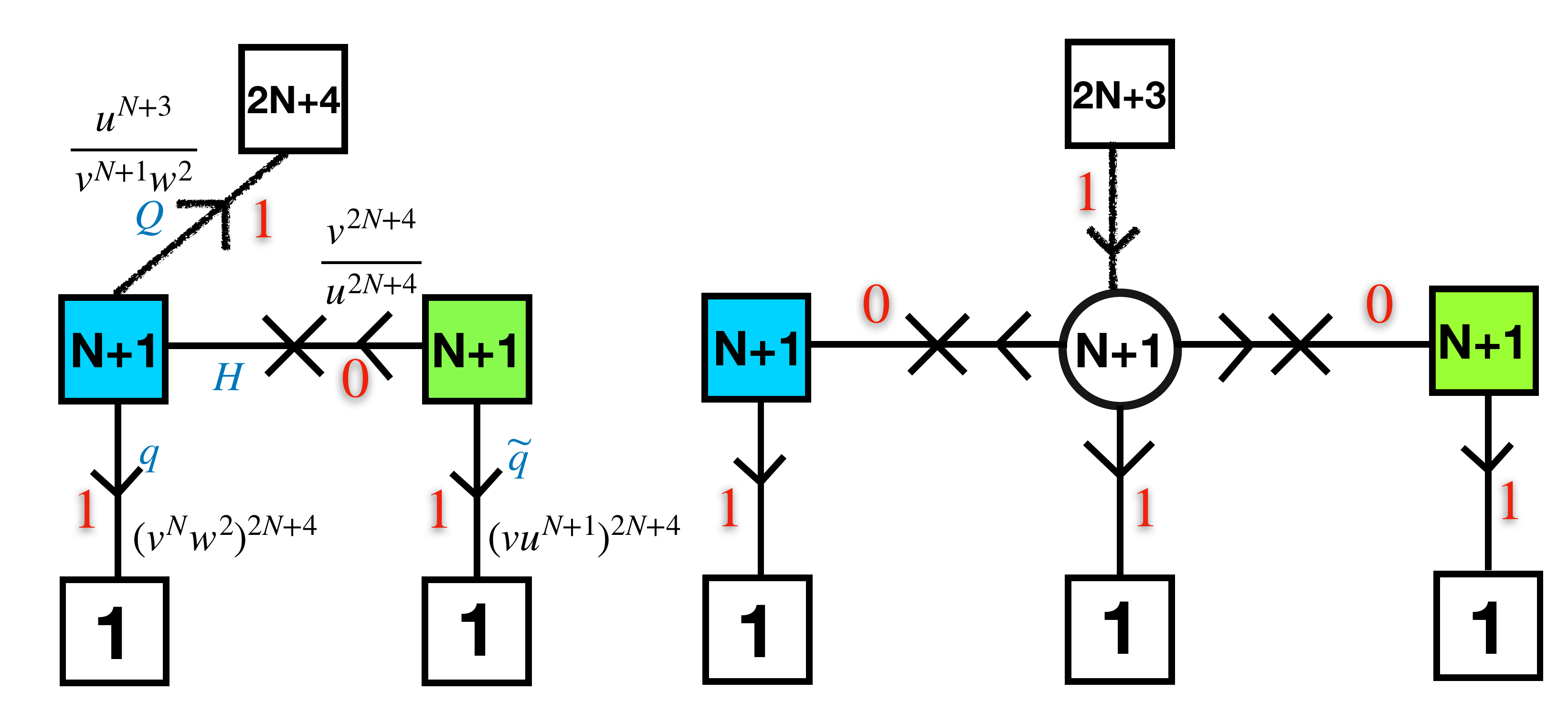}
    \caption{The two punctured spheres obtained by closing the minimal puncture. On the left we have the result of closing with the baryonic vev. The moment maps of the left puncture are the fields $Q$ and $q$ as well as the baryonic operator $H^N\widetilde q$. The moment maps for the puncture on the right are the field $\widetilde q$, the mesonic operators $QH$, and the baryon $H^Nq$. On the right we have the result of closing the minimal puncture with a mesonic vev. The crosses flip the baryons built from each of the bi-fundamental fields. }
    \label{F:DN3tubes}
\end{figure}

\begin{figure}[htbp]
	\centering
  	\includegraphics[scale=0.34]{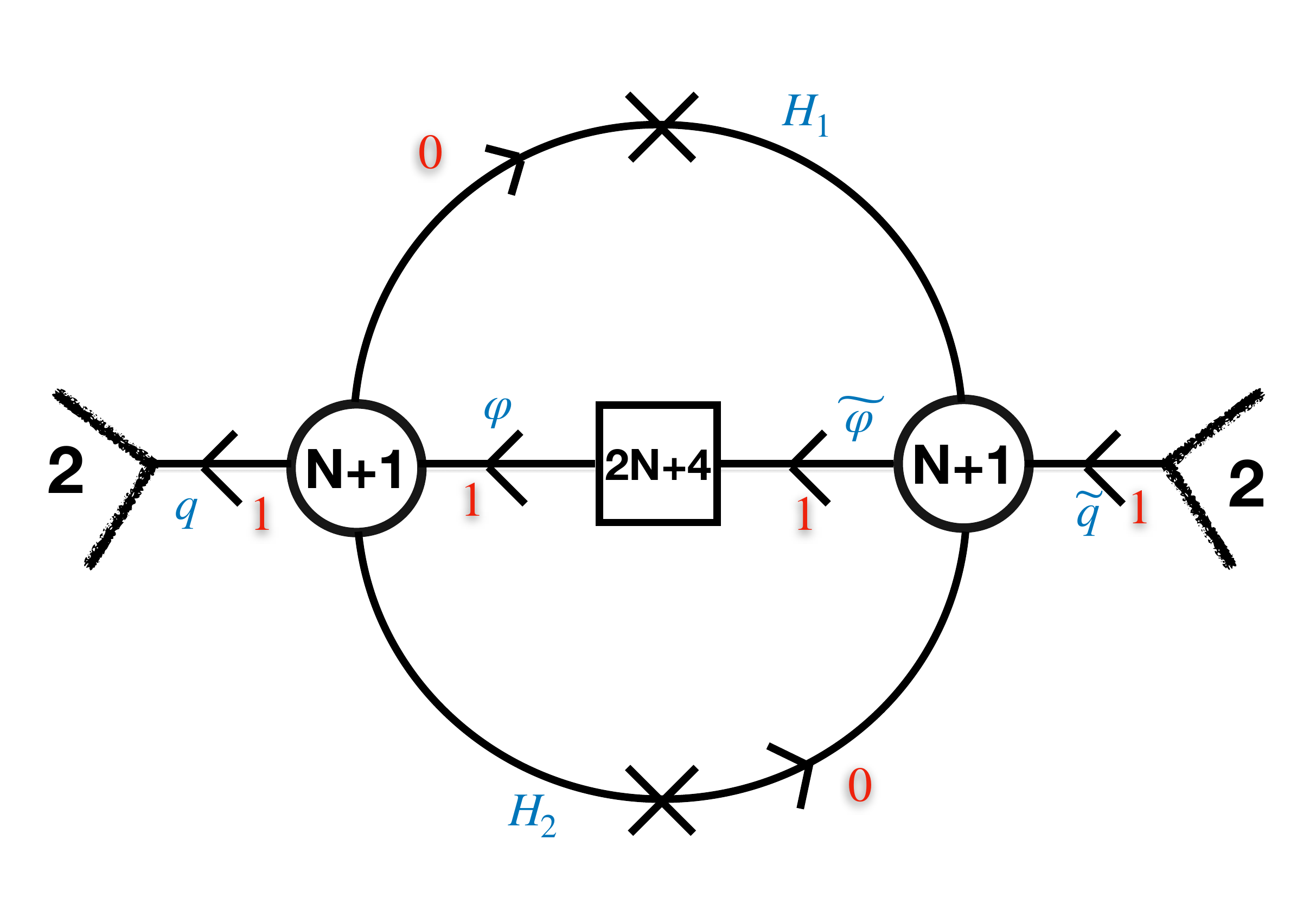}
    \caption{The torus theory with one unit of flux breaking $SO(4N+12)$ to $SO(4N+8)\times SU(2)\times U(1)$. In addition to the superpotentials corresponding to the triangles and the flip fields (denoted by crosses) we also have superpotential terms  of the form $qH_i^N\widetilde q$. The open squares denote that the two $SU(2)$ groups associated to them are identified. }
    \label{F:Dtorus}
\end{figure}

Let us now close the puncture with a vev to the baryonic moment map to obtain a flux of
$${\cal F}+\Delta{\cal F} = (1,-1,0,0,0,0,0,\cdots)\,.$$ This is again half a unit of $SO(4N+8)\times SU(2)\times U(1)$ flux related to the one we started with by an action of the Weyl group of $SO(4N+12)$.
On the other hand if we give a vev to one of the mesonic moment maps (the first for example), we obtain,
$${\cal F}+\Delta{\cal F} = (-1,-1,2,0,0,0,\cdots)\,,$$   which is the minimal flux breaking $SO(4N+12)$ to $SO(4N+6)\times SU(2)\times U(1)\times U(1)$.
We can compute the RG flow following the vevs and the resulting theories in both cases are depicted in Figure \ref{F:DN3tubes}. We can $\Phi$-glue together these tubes to obtain theories corresponding to tori with flux but no punctures. For example $\Phi$-gluing two copies of the  $SO(4N+8)\times SU(2)\times U(1)$ tube to a torus one obtains the quiver of Figure \ref{F:Dtorus}.
Note that a model corresponding to a torus with this amount of flux was constructed in \cite{Kim:2018bpg} using a completely different method. There tubes with two maximal punctures of $SU(N+1)$ and $USp(2N)$ were obtained. The quiver in Figure \ref{F:Dtorus} was exactly derived by gluing two such tubes and using the Intriligator-Pouliot duality \cite{Intriligator:1995ne} (see Figure 7 in \cite{Kim:2018bpg}). This is a non trivial check of our claims. In particular the checks of symmetry enhancements and anomlaies were already performed in \cite{Kim:2018bpg}.

Let us note that we can also flip the moment maps to obtain other types of two punctured spheres with flux. Moreover, we can close also the maximal punctures. Closing maximal punctures by a single vev will reduce the symmetry of the puncture but not completely close it. Performing a sequence of such flows will lead to a three punctured sphere with a maximal, a minimal and a more generic puncture at each step of the sequence. In particular flipping some of the vevs and closing the maximal punctures we will keep the $SU(N+2)$ gauge group while reducing the number of flavors. Thus for $N>1$ the $SU(N+2)$ SQCD with $N_f<2N+4$ will be related to certain trinions with generic types of punctures.

Finally we want to make two important comments. The first comment is as follows. 
Note that the theory of Figure \ref{F:Dtorus} has a similar structure to the theory in Figure \ref{F:wheelD}. This can be seen by taking $N\to N-1$ in the latter, and removing the flip fields flipping the baryons of the bi-fundamentals and the order $N+2$ superpotential locking the $SU(2)$ symmetries together in the former. 
 Thus, we see that the torus with $k$ units of flux for $D_{N+4}$ conformal matter is the same,
 up to gauge singlet fields  and superpotential interactions, as a torus with zero flux and $2k$ minimal punctures for $D_{N+3}$ minimal conformal matter. This is the general observation of \cite{Razamat:2019mdt}: the reduction of a theory from $6d$ on a Riemann surface with some value of flux can be related to a reduction of a theory obtained by a $6d$ Higgs branch flow on a surface with the same genus but possibly additional number of punctures. The number of additional punctures is determined by flux. This is the procedure which was used in \cite{Razamat:2019ukg} to obtain three punctured spheres of $D_{N+3}$ minimal conformal matter starting from tori with flux of $D_{N+4}$ matter built from tubes of \cite{Kim:2018lfo}. These tubes have $SU(2)^{N+1}$ type maximal punctures.
 Here we relate tori built from tubes with $SU(N+2)$  type maximal punctures of \cite{Kim:2018bpg}, which suggests that the trinions we obtain here should be derivable by the same procedure as in \cite{Razamat:2019ukg} but using the description of \cite{Kim:2018bpg} (particularly, the vev is given to the flip fields denoted by crosses in Figure \ref{F:Dtorus}). The equality of various models corresponding to higher genus surfaces derived in this paper and in \cite{Razamat:2019ukg} then would be related to the dualities between tori theories with flux discussed in Appendix B of \cite{Kim:2018bpg}.

\begin{figure}[htbp]
	\centering
  	\includegraphics[scale=0.38]{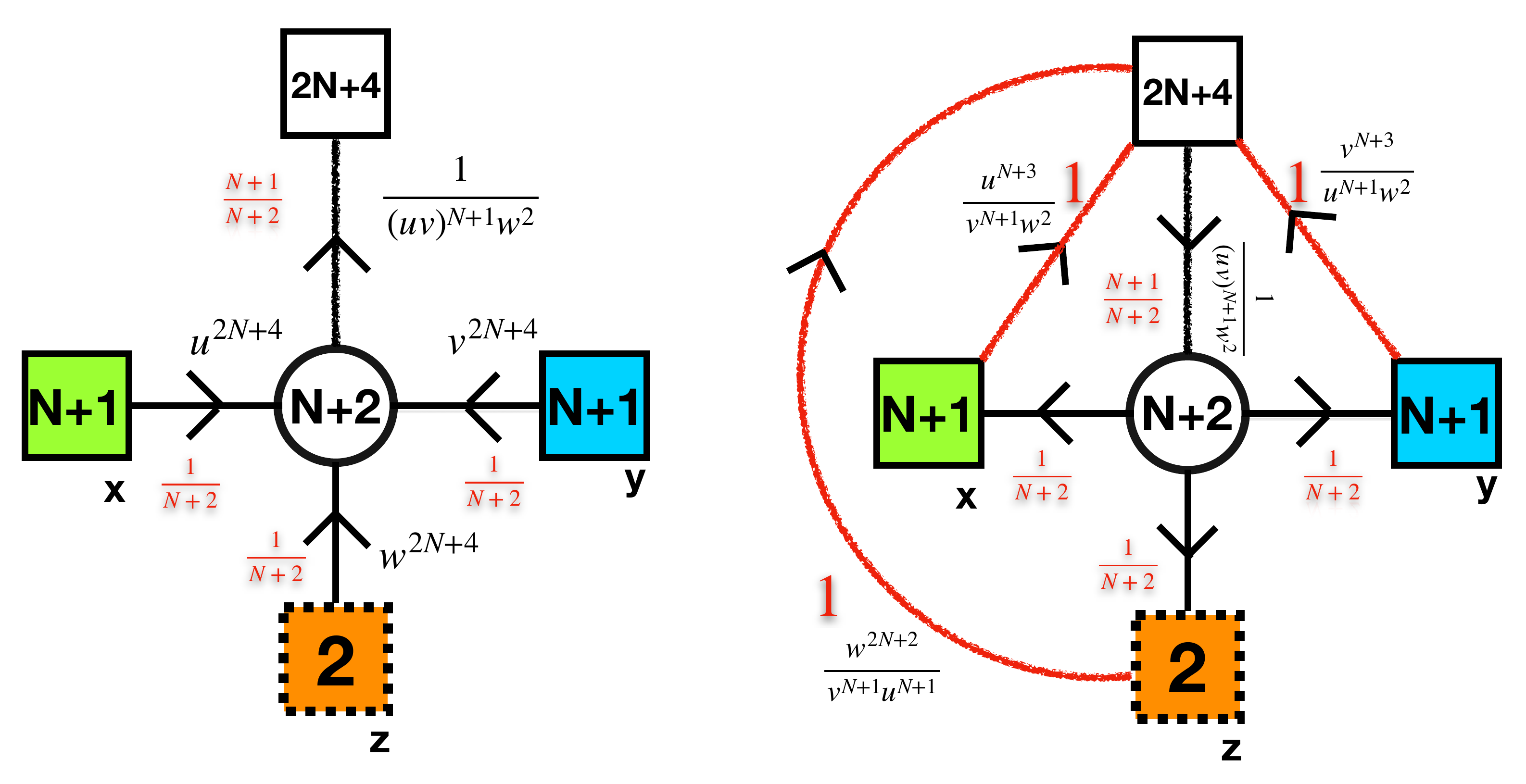}
    \caption{Seiberg duality of the trinion. On the left is the original trinion and on the right the Seiberg dual one. The baryonic symmetry stays intact while all the other symmetries are conjugated. Geometrically we flip the signs of the flux components corresponding to mesonic moment maps and exchange the two baryonic components. As the flux is zero for the mesonic components and equal for the two baryonic ones, its value does not change. The conjugation of charges of the mesonic moment maps changes the punctures type. However, flipping them we obtain the same punctures as on the left. Then, as the geometric data of the theory does not change under these manipulations, we expect to get dual models.}
    \label{F:seiberg}
\end{figure}

The second comment  we want to emphasise  is that
the understanding of $SU(N+2)$ SQCD in the middle of the conformal window as a three punctured sphere with flux \eqref{trinflux} gives us a geometric way to interpret its Seiberg duality \cite{Seiberg:1994pq}.  The Seiberg dual is given by the same gauge theory but with the mesons flipped. Lets us understand this in our language. Note that the baryonic symmetry of the trinion in Figure \ref{F:trinionD} is $1/(uv)^{N+1}w^2$. This symmetry stays intact under Seiberg duality. The $SU(2N+4)\times SU(2N+4)$ representations on the other hand are conjugated. One of the two $SU(2N+4)$ appears in our quiver description of the theory explicitly while the other is split into $SU(N+1)^2\times SU(2) \times U(1)^2$. The latter two $U(1)$ symmetries are parameterized by $u^{N+3}/(w^2 v^{N+1})$ and $v^{N+3}/(w^2 u^{N+1})$. Seiberg duality conjugates these two symmetries. Note that these are precisely the symmetries under which the mesonic moment map operators $M_u$ and $M_v$ \eqref{momentmapsN} are charged, and the mesonic component of $M_w$ is charged under the product of these symmetries. 
On the other hand the baryonic components have charges,
\be
&M_u:\;& {{\bf N+1}}^x\otimes{\bf 1}_{(u v^{N+1})^{2N+4}}\oplus
\overline{{\bf N+1}}^x\;\otimes {\bf 1}_{(u^Nw^2)^{2N+4}}\to \\
&&
\overline{{\bf N+1}}^x\;\otimes {\bf 1}_{(u^Nw^2)^{2N+4}}\oplus{{\bf N+1}}^x\otimes{\bf 1}_{(u v^{N+1})^{2N+4}}\nonumber\\
&M_v:\;& {{\bf N+1}}^y\otimes{\bf 1}_{(v u^{N+1})^{2N+4}}\oplus
\overline{{\bf N+1}}^y\;\otimes {\bf 1}_{(v^Nw^2)^{2N+4}}\to \nonumber\\
&&
\overline{{\bf N+1}}^y\;\otimes {\bf 1}_{(v^Nw^2)^{2N+4}}\oplus{{\bf N+1}}^y\otimes{\bf 1}_{(v u^{N+1})^{2N+4}}\nonumber\\
&M_w:\;&{\bf 2}^z\otimes\left({\bf 1}_{(w v^{N+1})^{2N+4}}\oplus {\bf 1}_{(w u^{N+1})^{2N+4}}\right)\to {\bf 2}^z\otimes \left({\bf 1}_{(w u^{N+1})^{2N+4}}\oplus{\bf 1}_{(w v^{N+1})^{2N+4}}\right)\,.\nonumber
\ee 
That is, the conjugation of Seiberg duality simply exchanges the baryonic moment maps.
Note that the flux \eqref{trinflux} has zeros in all the mesonic components of the moment map symmetries and the fluxes of the two baryonic components are equal, and thus the Seiberg dual theory should be associated to a theory with the same value of flux albeit with the mesonic components of the punctures having conjugate charges. The mesonic flip fields added in the duality transform these components to have the same charges as in the original frame. Thus geometrically the two frames correspond to a surface with same flux and same punctures and thus have to be the same, as indeed Seiberg duality claims. See Figure \ref{F:seiberg} for illustration. In other words the action of Seiberg duality can be thought of group theoretically as an element of the Weyl symmetry group of $SO(4N+12)$ acting on the flux vector permuting first two components  and flipping the signs of the rest. The flux vector associated to the trinion  is invariant under this operation but the type of punctures changes, which is fixed by adding the flip fields.  Permutation of the last $2N+4$ components  is also an  operation keeping the theory invariant, but it also does not change the types of punctures and thus it is a symmetry rather than a duality. For earlier very related interplays between the action of the Weyl group and dualities see \cite{Spiridonov:2008zr,Dimofte:2012pd}.

\section{Discussion}\label{sec:A}

In this paper we have suggested a geometric interpretation of $SU(M)$ SQCD in the middle of the conformal window, that is $N_f=2M$. These models can be obtained as 
compactifications of $D_{M+1}$ minimal conformal matter in $6d$ on three punctured spheres with two maximal $SU(M-1)$ punctures and one minimal $SU(2)$ puncture. For the case of $M=3$ the minimal and maximal punctures coincide. The discussion here can be deepened and extended in various ways. Let us discuss some of them.

First, the models constructed from the trinions have interesting dynamics. The superpotentials and gaugings in some of the compactifications are irrelevant and can be shown to be dangerously irrelevant for a subset of these cases. It will be very interesting to have a general statement about  the IR physics of these models: whether we always (and if not always in which cases) end up with an interacting SCFTs in the IR. In other compactifications, such as $N$ M5 branes on $A_{k-1}$ singularities \cite{Gaiotto:2015usa}, it has been shown that on tori for many cases some of the superpotentials are irrelevant and lead to decoupled chiral fields in the IR \cite{Bah:2017gph}. These superpotentials are found crucial in cases where symmetry is enhanced in the IR, and serendipitously in those cases they are relevant, and also are needed to match anomalies with $6d$. The irrelevant supeprotentials take in such cases the form of flipping baryonic operators. In our examples we have similar types of interactions, but instead the baryons we consider transform in a non trivial representation of the gauge group. 
It seems that to understand explicitly the QFTs arising as general compactifications of $6d$ models there is a need to get a better grip over complicated dynamics.  For example, in previous works such compactifications were understood in terms of field theoretic constructions were one gauges a symmetry which is emergent in the IR \cite{Gadde:2015xta,Razamat:2016dpl,Kim:2017toz,Pasquetti:2019hxf,Razamat:2019ukg,Agarwal:2018ejn}. Here, at least in the E-string case, all our gauge symmetries are visible in the Lagrangian but dynamics is not trivial. It will be very interesting to understand these issues further. In the case of $N=M-2>1$ in order to construct models corresponding to higher genus we needed to gauge symmetries conjecturally emerging in the IR.

Another point worth stressing and which should be studied further is the interplay between group theory and dualities. We have observed in this paper, as well as in previous works, that some of the known and more novel IR dualities follow in geometric constructions from group theory arguments. These ultimately stem from the claim that choices of flux related by Weyl symmetry of the group and different decompostion of the flux should give rise to equivalent theories. This relation should be further investigated. Let us mention that some known dualities have been observed in the past to form orbits of the action of the Weyl symmetry group \cite{Spiridonov:2008zr,Dimofte:2012pd,Razamat:2017hda,Amariti:2018wht,Benvenuti:2018bav}, but not all of these have at the moment a geometric interpretation (though some do). Let us also mention that it will be interesting to relate the dualities discussed here to some of the  class ${\cal S}$ constructions, {\it e.g.} \cite{Gadde:2013fma}. 

A very interesting question is whether any of the understandings we have developed here can be applied to other compactification setups, see {\it e.g.}   \cite{Kim:2018lfo,Chen:2019njf,Razamat:2018gbu,Sela:2019nqa,Zafrir:2018hkr,Zafrir:2019hps} and for a review of $6d$ SCFTs see \cite{Heckman:2018jxk}. In many of these we understand, maybe partially, how to build theories corresponding to two punctured spheres and it would be very important to derive a general understanding of compactifications on three punctured spheres which would lead to constructions of theories corresponding to arbitrary surfaces.

There are numerous geometric constructions of various supersymmetric quantum field theories in four dimensions utilizing string and M/F-theory methods, see {\it e.g} \cite{Elitzur:1997fh,Giveon:1998sr,Bershadsky:1996gx,Cachazo:2001sg,Franco:2005rj,Garcia-Etxebarria:2015wns,Apruzzi:2018oge}. Many of the different constructions result in related theories in $4d$. For example: brane constructions of  \cite{Elitzur:1997fh} give rise to the SQCD models we obtained here; toric quivers of \cite{Franco:2005rj,Franco:2005sm} related to various CY geometries  are closely related to models coming from M5 branes probing $A$ type singularity compactified on tori with flux \cite{Bah:2017gph}; theories coming from M5 branes probing a D-type singularity compactified on tori with flux are related to M5 branes on $A$ type singularities compactified on spheres with punctures \cite{Kim:2018bpg}; M5 branes probing $ADE$ type singularities compactified on tori with no flux are related to class ${\cal S}$ theories \cite{Gaiotto:2009we,Gaiotto:2009hg},  M5 branes compactified on spheres with punctures \cite{Ohmori:2015pia}; a trinion for the E-string theory can be constructed as a deformation of a trinion of two M5 branes probing an $A_1$ singularity \cite{Kim:2017toz}; and many more. Developing a better understanding of such interrelations would be a very interesting research direction to pursue.

Let us also mention here that in the general framework of interconnections of supersymmetric indices of $4d$  theories obtained in compactificaations on Riemann surfaces to TQFTs \cite{Gadde:2009kb} and integrable quantum mechanical models \cite{Gaiotto:2012xa,Gadde:2011uv}, the trinion for the E-string we obtain here should be a Kernel function  \cite{MR3057192,Nazzal:2018brc,MR3313680} of the $BC_1$ van Diejen model \cite{MR1275485}. The higher $N$ trinions should give rise to generalizations of this model, properties of which would be very interesting to explore.

\section*{Acknowledgments}

We are grateful to Gabi Zafrir for interesting comments on the draft of the paper.
This research is supported in part by Israel Science Foundation under grant no. 2289/18, by I-CORE  Program of the Planning and Budgeting Committee, by a Grant No. I-1515-303./2019 from the GIF, the German-Israeli Foundation for Scientific Research and Development, and by BSF grant no. 2018204.

\appendix

\section{The trinion of \cite{Razamat:2019ukg} and ${\cal N}=1$ $SU(3)$ $N_f=6$ SQCD }\label{app:oldtrinion}

\begin{figure}[htbp]
	\centering
  	\includegraphics[scale=0.36]{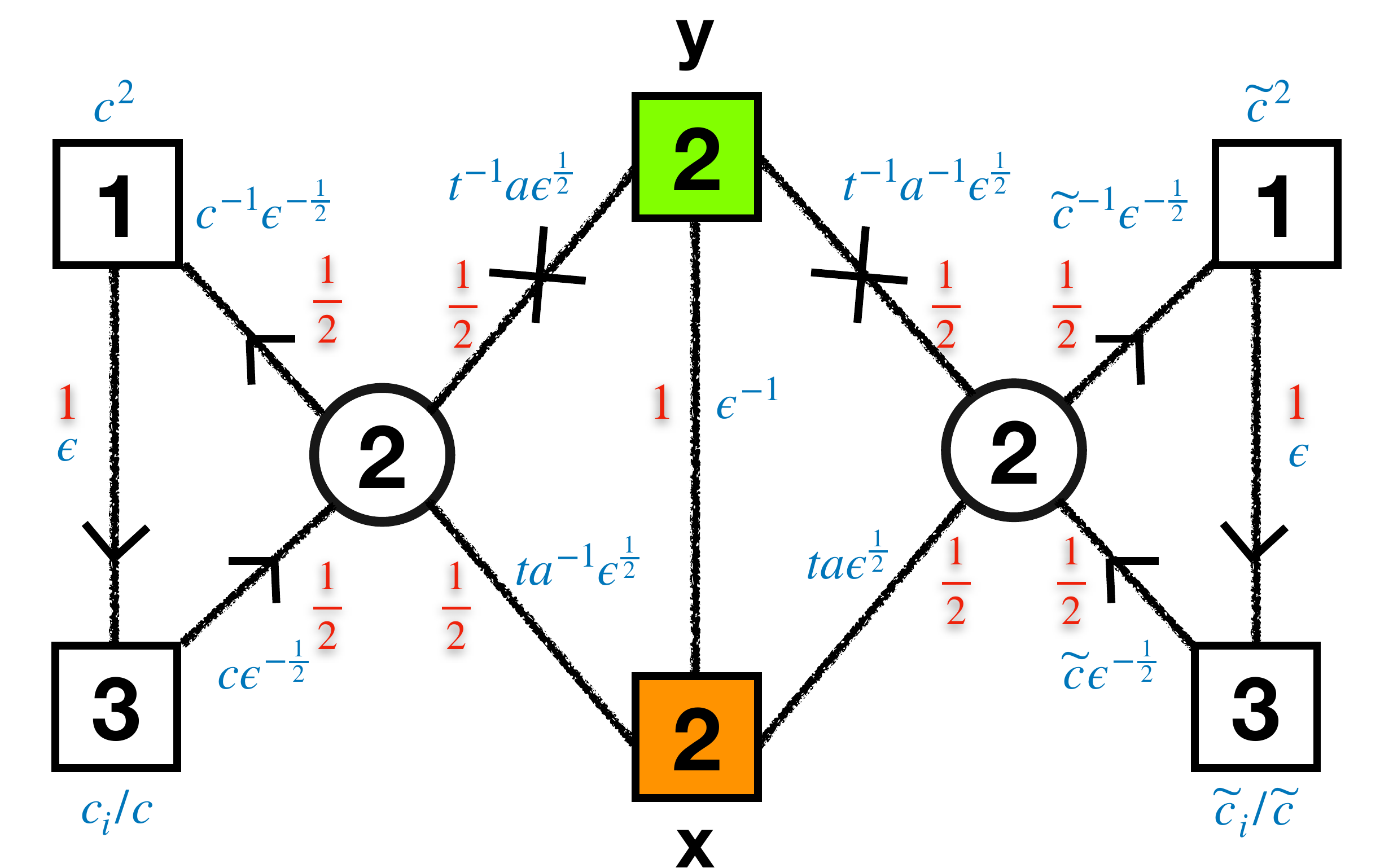}
    \caption{The E-string trinion found in \cite{Razamat:2019ukg} with two $SU(2)$ punctures seen in the UV and another puncture with enhanced symmetry $SU(2)_\epsilon$ in the IR. This trinion has vanishing flux. We use the fugacity definitions $c=(c_1 c_2 c_3)^{1/3}$ and $\widetilde{c}=(\widetilde{c}_1 \widetilde{c}_2 \widetilde{c}_3)^{1/3}$ to simplify the quiver. Superpotentials corresponding to the triangles and the flip fields (denoted by crosses) are turned on. }
    \label{F:OldTrinion}
\end{figure}

In this appendix we will relate the trinion obtained in this paper to the one found in \cite{Razamat:2019ukg}. For simplicity we will only discuss the trinions of the E-string. The  E-string trinion of \cite{Razamat:2019ukg}  is a gauge theory of two $SU(2)$ gauge nodes as shown in Figure \ref{F:OldTrinion}. The map between the two constructions will serve as another test for the validity of our results.

\begin{figure}[htbp]
	\centering
  	\includegraphics[scale=0.32]{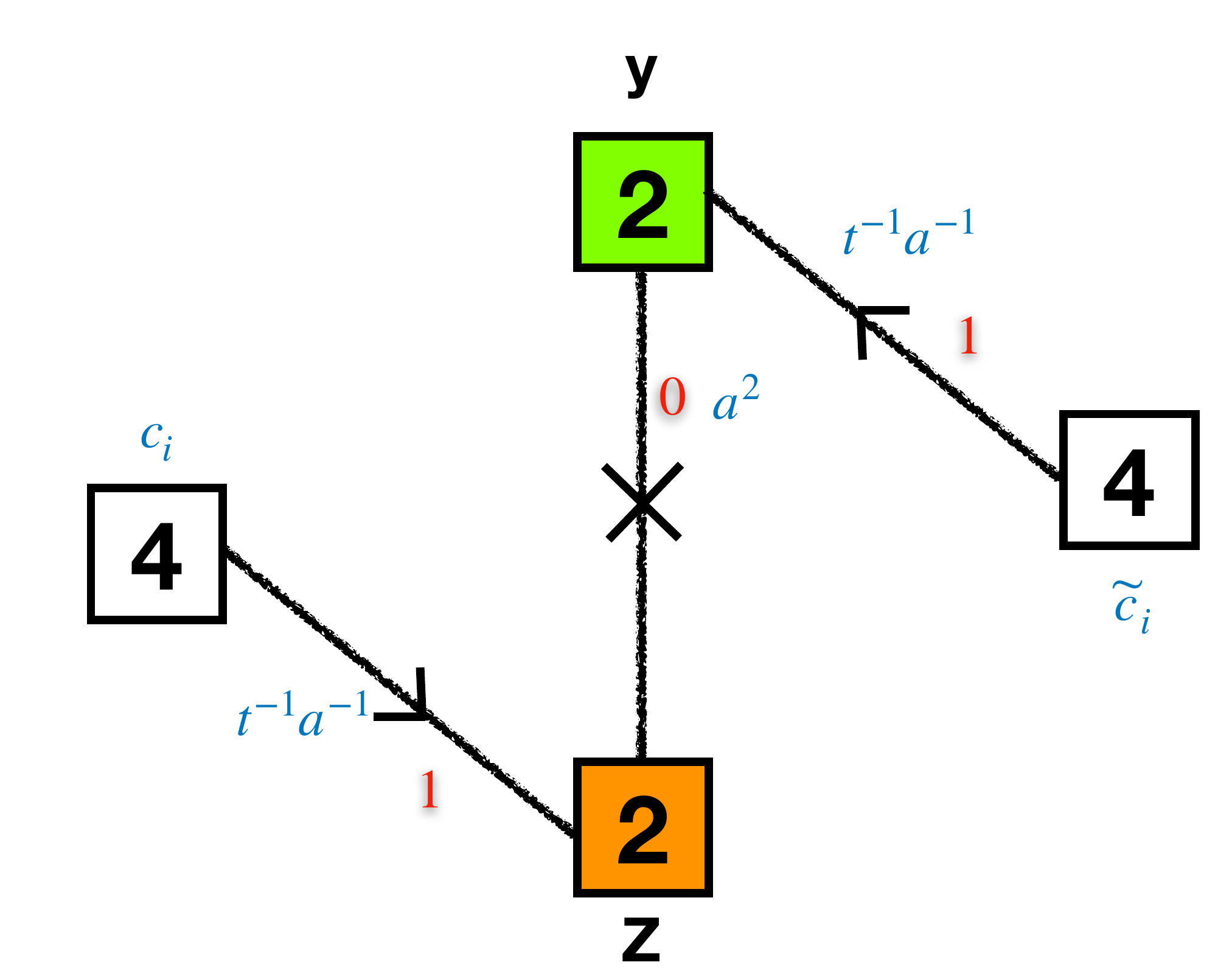}
    \caption{E-string $a$ flux tube. Here we use the fugacity definitions $c_4=(c_1 c_2 c_3)^{-1}$ and $\widetilde{c}_4=(\widetilde{c}_1 \widetilde{c}_2 \widetilde{c}_3)^{-1}$. A superpotential corresponding to the flip field (denoted by a cross) is turned on. }
    \label{F:Old2NewFluxTube}
\end{figure}

We will compare between spheres with four punctures and vanishing flux built in two ways. The first will be constructed using the trinion of \cite{Razamat:2019ukg}  $\Phi$-glued through the $y$ puncture to the corresponding $y$ puncture in the flux tube shown in Figure \ref{F:Old2NewFluxTube}. This will give a trinion of non vanishing flux. In addition we need to partially flip the $x$ and $z$ punctures by adding a singlet field of R-charge 1 and charges marked by the fugacities $t^{-1} a^{-1} \widetilde{c}_3^{-1}$ in the fundamental of $SU(2)_x$ and $t a^{-1} \widetilde{c}_3^{-1}$ in the fundamental of $SU(2)_z$. Finally we will take two copies of this trinion and $S$-glue them through the $\epsilon$ puncture to  generate a four punctured sphere. The second construction is of two  trinions of this paper $S$-glued through the $w$ puncture\footnote{Meaning the puncture that has a field charged under its $SU(2)$ symmetry and the $U(1)_w$ symmetry.} to one another to generate the four puncture sphere.

In order to fully identify the theories we need to identify how to map the symmetries, or equivalently the  fugacities for the symmetries. By comparing the $x$ moment maps in the  trinion of \cite{Razamat:2019ukg} described in Figure \ref{F:OldTrinion} with the $u$ puncture moment maps of the  trinion of this paper shown in the beginning of Section \ref{sec:E}, we find that,
\be
ta^{-1}c_{i}=a_{i}u^{4}\left(vw\right)^{-2}\ &,& \quad ta\widetilde{c}_{1}=a_{5}u^{4}\left(vw\right)^{-2}\ ,\quad ta\widetilde{c}_{2}=a_{6}u^{4}\left(vw\right)^{-2} \nonumber\\
t^{-1}a^{-1}\widetilde{c}_{3}^{-1}=u^{6}v^{12}\ &,&\quad ta\widetilde{c}_{4}=u^{6}w^{12}\,.
\ee
Under the mapping described one finds the two theories match.

\begin{figure}[htbp]
	\centering
	\includegraphics[scale=0.42]{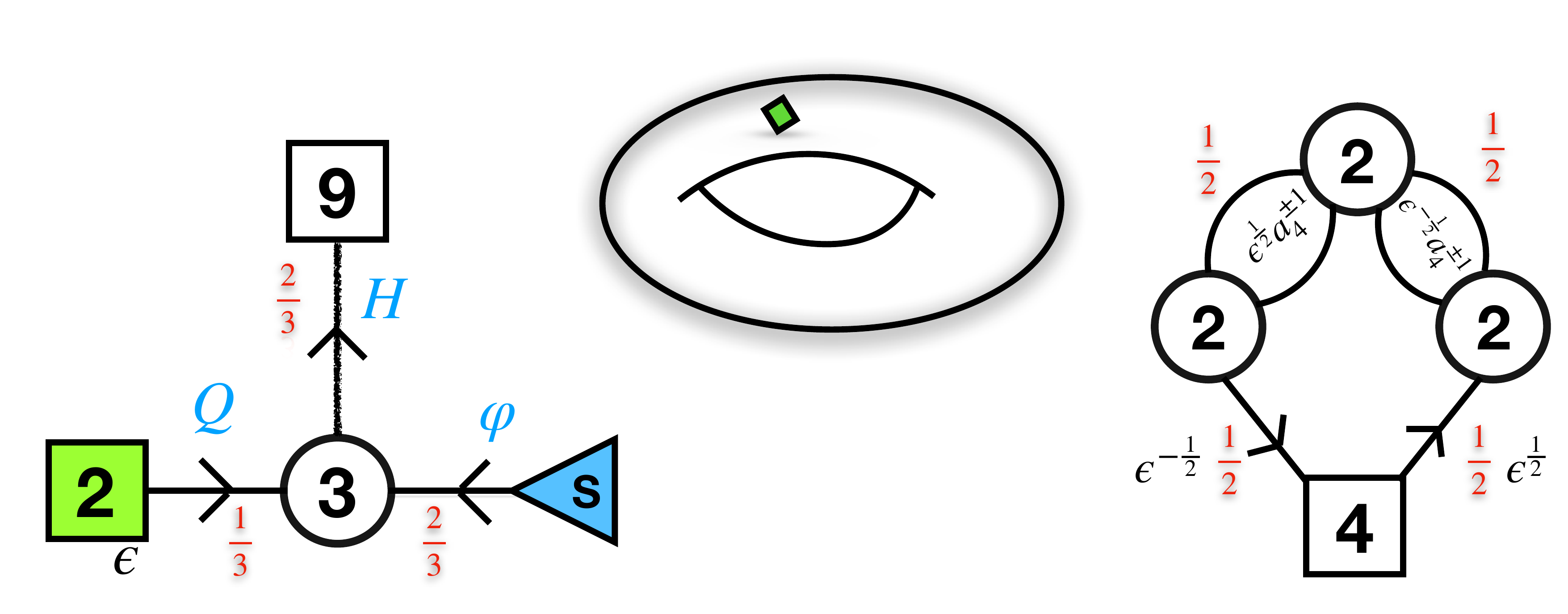}
    \caption{A duality following from gluing two punctures of the trinion together. On the left a torus built from the trinion presented in this paper and on the right one built from the trinion of \cite{Razamat:2019ukg}. Both models are of a torus with one puncture.
    The puncture symmetry is $SU(2)_\epsilon$ and on the right only the Cartan is visible in the UV.  As we glue two different types of punctures together the global symmetry is broken to $SO(9)$. We have superpotentials consistent with the charges in the Figure. We parametrize the Cartan of $SO(9)$ by  four fugacities $a_i$.
     The $SU(4)$ on the right is parametrized by $\sqrt{a_1a_2/a_3}$, $\sqrt{a_1a_3/a_2}$, and $\sqrt{a_3a_2/a_1}$.  }
    \label{F:simpleduality}
\end{figure}

One can generate various types of dualities using the two descriptions and let us mention here one example.
Let us S-glue two punctures of the same trinion together. The resulting theory is depicted on the left side of Figure \ref{F:simpleduality}. The $SU(2)$ gauging of the two punctures of the trinion presented in this paper has $N_f=3$ and is given in the IR by a collections of gauge singlets. These will provide two anti-fundamentals, two index symmetric, and two index antisymmetric: The last is just another anti-fundamental and the second is a six dimensional representation.
The punctures of the trinion are of different types, colors, and thus gluing them will necessarily break some of the symmetry. The only symmetry preserved by the gluing is a subgroup of $SU(6)$ and $U(1)_{v/w}$, where the fundamental fields transforming under the gauged $SU(2)$ are charged under $U(1)_{v,w}$. In fact the preserved  symmetries combine into $SO(9)$. The quiver is on the left side of Figure \ref{F:simpleduality}. In particular the symmetry $uvw$, in which the trinion has flux, is broken and thus the corresponding model should correspond to a torus with one puncture and zero flux (possibly with some discrete twists and/or fluxes \cite{Kim:2017toz,Bah:2017gph}; we leave investigations of these issues to future work). On the other hand we can start from the trinion of \cite{Razamat:2019ukg}, which has zero flux, and S-glue two punctures together to obtain a torus with one puncture (and possibly discrete twists). The dual model is depicted in Figure \ref{F:simpleduality} where we glued the two punctures that have an $SU(2)$ symmetry visible in the UV and slightly manipulated the quiver using the duality of \cite{Csaki:1997cu} on one of the trinion's $SU(2)$ nodes. 

The punctures of the two models are different and thus the discrete subtleties might give rise to differences in the model. Nevertheless, in this case (this is not necessarily true for higher values of $N$) there is a deformation we can perform on the right quiver to get the left one:
A relevant deformation that will break some symmetries and erase some of the data associated to them, such as fluxes \cite{Razamat:2016dpl}. The superpotentials turned on are quartic ones consistent with the symmetries denoted in Figure \ref{F:simpleduality}. Note that here the $U(1)_\epsilon$ symmetry enhances to $SU(2)_\epsilon$ on some locus of the conformal manifold as well as $U(1)_{a_i}$ symmetries enhance to $SO(9)$.\footnote{Note that this model can be obtained as an exactly marginal deformation of the compactification on a sphere with six punctures and zero flux of $A_1$ $(2,0)$ theory \cite{Benini:2009mz,Bah:2012dg}. In that context the enhancement of $U(1)_\epsilon$ to $SU(2)_\epsilon$ is due to the fact that it is the Cartan of the $SU(2)$  part of the $6d$ R-symmetry and as we have zero flux it becomes a global $SU(2)$ symmetry in $4d$.}
This is a rather simple but non trivial duality that follows from geometric considerations. One can indeed verify the anomalies of the two models as well as supersymmetric indices match.

The dynamics on the right hand side of the duality is rather simple: one can start with gauging the symmetries, which are just $3$ copies of asymptotically free $SU(2)$ $N_f=4$ gauge nodes, and then turn on quartic interactions which are exactly marginal deformations. On the  left hand side the dynamics is more involved so let us discuss it in more detail.
First, in addition to the R-symmetry depicted in the Figure, at the free point we have two anomaly free $U(1)$ symmetries under which the various fields have the following charges,
\begin{center}
\begin{tabular}{ |c|c|c|c|c| } 
 \hline
  & $U(1)_a$ & $U(1)_b$ \\ 
  \hline\hline
 $H$ & $1$ & $0$ \\ 
 $\varphi$&$0$&$1$\\
 $Q$ & $-\frac{9}2$ &$-\frac52$\\ 
 \hline
\end{tabular}
\end{center}
Then we can turn on cubic interactions of the form $H^2\varphi$ which are relevant. Doing so we break the $U(1)_b$ symmetry locking it onto $-2$ the $U(1)_a$. This deformation also breaks the $SU(9)$ of the quiver to $SO(9)$.
There are no operators violating the unitarity bounds. With this deformation the superpotential of the form $\det \varphi \sim \varphi^{3}$ is relevant. By turning it on we also break $U(1)_a$ and the six dimensional R-symmetry becomes the superconformal one. The theory thus flows to an SCFT and we claim that it is dual to the one on the right in Figure \ref{F:simpleduality}. One can take two copies of the theory corresponding to the one punctured torus and glue them together to form a theory corresponding to genus a two surface. This should be dual to the $g=2$ model of Figure \ref{F:wheel}. In fact this duality was already discussed in \cite{Razamat:2019vfd} and corresponds to the two different pair of pants decompositions of the genus two surface.



\bibliographystyle{ytphys}
\bibliography{refs}

\end{document}